\documentclass[]{fairmeta}

\title{
    \resizebox{\linewidth}{!}{How Good is Post-Hoc Watermarking With Language Model Rephrasing?}
} 

\author[\star, 1]{Pierre Fernandez}
\author[\star, 1]{Tom Sander}
\author[1]{Hady Elsahar}
\author[1]{Hongyan Chang}
\author[1]{Tom\'{a}\v{s} Sou\v{c}ek}
\author[1]{Valeriu Lacatusu}
\author[1]{Tuan Tran}
\author[1]{Sylvestre-Alvise Rebuffi}
\author[1]{Alexandre Mourachko}
  
\affiliation[1]{FAIR, Meta Superintelligence Labs}

\contribution[\star]{Core \& Equal contributors.}

\abstract{
Generation-time text watermarking embeds statistical signals into text 
for traceability of AI-generated content. 
We explore \emph{post-hoc} watermarking where an LLM rewrites existing text while applying generation-time watermarking, to protect copyrighted documents, or detect their use in training or RAG via watermark radioactivity.
Unlike generation-time approaches, which are constrained by how LLMs are served, this setting offers additional degrees of freedom for both generation and detection.
We investigate how allocating compute (through larger rephrasing models, beam search, multi-candidate generation, or entropy filtering at detection) affects the quality-detectability trade-off. 
Our strategies achieve strong detectability and semantic fidelity on open-ended text such as books.
Among our findings, the simple Gumbel-max scheme surprisingly outperforms more recent alternatives under nucleus sampling, and most methods benefit significantly from beam search.
However, most approaches struggle when watermarking verifiable text such as code, where we counterintuitively find that smaller models outperform larger ones. 
This study reveals both the potential and limitations of post-hoc watermarking, laying groundwork for practical applications and future research.

} 
 
\correspondence{\email{tomsander@meta.com}, \email{pfz@meta.com}}

\metadata[Code]{\url{https://github.com/facebookresearch/textseal}}

\PassOptionsToPackage{numbers, sort&compress}{natbib}

\usepackage[utf8]{inputenc} %
\usepackage[T1]{fontenc}    %
\usepackage{url}            %
\usepackage[rightcaption]{sidecap}
\usepackage{amsfonts}       %
\usepackage{nicefrac}       %
\usepackage{microtype}      %
\usepackage{xcolor}         %
\usepackage{enumitem}
\usepackage{nicefrac}       %

\usepackage{amsmath}        %
\usepackage{dsfont}        %

\usepackage{xltabular}
\usepackage{graphicx}       %
\usepackage{caption} 
\usepackage{siunitx} 
\usepackage{booktabs}
\usepackage{multirow} 
\usepackage{tabularx}
\usepackage{subcaption}
\usepackage{wrapfig}
\usepackage{tabulary} 
\usepackage{bbm}
\usepackage{lipsum}

\usepackage[most]{tcolorbox} 
\usepackage{listings}

\usepackage[utf8]{inputenc}
\usepackage{booktabs}
\usepackage{tabularx}
\usepackage[table]{xcolor}
\usepackage{geometry}
\usepackage{microtype} 
\usepackage{amsmath} %

\newcolumntype{Y}{>{\raggedright\arraybackslash}X}
\newcommand{\statsep}{\quad$\bullet$\quad}

\tcbuselibrary{skins, breakable, raster}
\usepackage{tikz}       %
\usetikzlibrary{decorations.pathreplacing, calc, shadows}

\usepackage{titlesec}
\titlespacing*{\paragraph}{0pt}{1em}{1em}

\usepackage{ifthen}
\newboolean{set_me_to_remove_todos}
\setboolean{set_me_to_remove_todos}{true} %
\newcommand{\todo}[1]{{\color{red} [\textbf{TODO}: #1]}}
\newcommand{\pierre}[1]{{\color{blue} [\textbf{Pierre}: #1]}}
\newcommand{\hongyan}[1]{{\color{orange} [\textbf{Hongyan}: #1]}}
\newcommand{\ismail}[1]{{\color{purple} [\textbf{Ismail}: #1]}}

\ifthenelse{\boolean{set_me_to_remove_todos}}{
    \renewcommand{\todo}[1]{}
    \renewcommand{\pierre}[1]{}
    \renewcommand{\hongyan}[1]{}
    \renewcommand{\ismail}[1]{}
    
}{}
\newcommand{\tnl}{\par\vspace{4pt}}  %

\definecolor{badred}{rgb}{0.8,0.2,0.2}
\definecolor{basecolor}{HTML}{E63946}
\definecolor{ftcolor}{HTML}{457B9D}
\definecolor{ftaugcolor}{HTML}{2A9D8F}
\definecolor{ftaugsynccolor}{HTML}{e2b33c}
\definecolor{textonlyorange}{HTML}{F4A261}

\usepackage{pifont}
\usepackage{xspace}
\newcommand{\cmark}{\ding{51}\xspace}%

\newcommand{\sk}{s}
\newcommand{\prob}[0]{\mathbb{P}}

\def\1{\mathbf{1}}

\def\V{\mathcal{V}}
\def\H{\mathcal{H}}

\usepackage{xspace} %

\newcommand{\eg}{e.g.,\@ }
\newcommand{\ie}{i.e.,\@ }

\newcolumntype{x}[2]{S[table-format=#1.#2,table-auto-round]}

\usepackage{tikz}

\usetikzlibrary{shapes.misc, positioning, decorations.pathreplacing}
\usetikzlibrary{positioning, backgrounds, fit, calc}

\definecolor{green1}{HTML}{008000}
\definecolor{green2}{HTML}{03ad55}
\definecolor{red1}{HTML}{FF0000}
\definecolor{red2}{HTML}{880000}

\lstdefinestyle{mystyle}{
    breaklines=true,
    basicstyle=\scriptsize\sffamily,
    numbers=none,
    language={},
    framextopmargin=0pt,
    framexbottommargin=0pt,
    breakindent=0pt,
    showspaces = false,
    keywordstyle=\bfseries,
    showstringspaces=false,
    columns=fullflexible,
    morekeywords={acoustic, guitar, thatch},
    moredelim=**[is][\color{green1}]{@}{@},
    moredelim=**[is][\color{red2}]{^}{^}
}

\newtcblisting{prompt}[1][]{
    arc=3pt, outer arc=3pt,
    width=.93\linewidth,
    left=1mm,
    top=0mm,
    bottom=0mm,
    title=\userheader, 
    colback=red!5!white,
    colframe=red!50!black,
    fonttitle=\bfseries,
    listing only, 
    listing options={style=mystyle,deletekeywords={}},
    breakable,
    #1
}

\definecolor{oai}{HTML}{10a37f}
\newtcblisting{chameleon}[2][]{
    arc=3pt, outer arc=3pt,
    width=.93\linewidth,
    left=1mm,
    top=0mm,
    bottom=0mm,
    title=\chameleonheader, 
    colback=oai!5!white,
    colframe=oai!40!black,
    fonttitle=\bfseries,
    listing only, 
    listing options={style=mystyle,deletekeywords={acoustic,guitar,thatch}},
    breakable,
    after upper={
      \vspace{0.5em}
      \includegraphics[width=0.45\textwidth]{prompts/#2.png}
    },
    #1
}
\newtcblisting{detector}[3][]{
    arc=3pt, outer arc=3pt,
    width=.93\linewidth,
    left=1mm,
    top=0mm,
    bottom=0mm,
    title=\detectorheader{#3}, 
    colback=metablue!5!white,
    colframe=metablue!40!black,
    fonttitle=\bfseries,
    listing only, 
    listing options={style=mystyle,deletekeywords={acoustic,guitar,thatch}},
    breakable,
    after upper={
      \vspace{0.5em}
      \includegraphics[width=0.45\textwidth]{prompts/#2.png}
    },
    #1
}

\newcommand{\chameleonheader}{
    \begin{tikzpicture}
      \node[anchor=north] {\pgftext{\includegraphics[width=0.34cm]{figures/assets/chameleon}}};
    \end{tikzpicture}
    ~Chameleon
}
\newcommand{\userheader}{
    \begin{tikzpicture}
      \node[anchor=north] {\pgftext{\includegraphics[width=0.34cm]{figures/assets/user2}}};
    \end{tikzpicture}
    ~User Prompt
}
\newcommand{\detectorheader}[1]{
    \begin{tikzpicture}
      \node[anchor=north] {\pgftext{\includegraphics[width=0.34cm]{figures/assets/drop}}};
    \end{tikzpicture}
    ~Watermark Detector (#1)
}

\begin{document}
\sisetup{text-series-to-math = true}

\maketitle

\section{Introduction}

Post-hoc text watermarking inserts an algorithmically-detectable signal into an existing text while remaining imperceptible to readers.
It serves several goals, such as copyright protection or traitor tracing.
Early approaches relied on hand-crafted modifications: synonym substitutions~\citep{topkara2006hiding, shirali2008new}, grammatical transformations~\citep{topkara2006words, topkara2006natural} or morphosyntactic alterations~\citep{meral2009natural}.
Recent methods based on deep neural networks follow a post-hoc paradigm where the input is the original text and the output is the watermarked text, using one watermark embedder and one watermark extractor~\citep{abdelnabi2021adversarial}, similarly to state-of-the-art post-hoc watermarking approaches for images, audio, and video. 
However, these methods are not effective, suffering from low capacity or unreliable detection, because they are highly constrained or easily broken by reversing the edits or synonyms.

With the emergence of large language models (LLMs), their popularization through ChatGPT~\citep{chatgpt2022}, and growing concerns about potential risks~\citep{crothers2022machine, weidinger2022taxonomy}, generation-time text watermarking algorithms have been introduced to help detect AI-generated text~\citep{aaronson2023watermarking,kirchenbauer2023watermark}. These algorithms have been deployed at scale, for instance, in Google's Gemini with SynthID~\citep{dathathri2024scalable}.
Most LLM watermarks alter the next token selection, for example, by promoting a specific set of tokens depending on previous tokens and a secret key.
Detection of the watermark in a text is then performed through a theoretically grounded statistical test that provides rigorous guarantees over the false positive rates. 
In addition, these methods achieve high watermark power while adding minimal latency by leveraging the entropy of the LLM to embed the signal.
However, in contrast with previous approaches, these methods require control at generation time and therefore cannot be applied to existing text.

A natural idea to extend these methods to post-hoc watermarking is to employ a paraphrasing LLM to re-generate the text, allowing the watermark to be injected at inference time without further model training.
This approach has been explored in data protection 
literature~\citep{jovanovic2025ward, sander2025detecting, rastogi2025stamp, 
zhang2025leave, lau2024waterfall}. 
For example, ``watermark radioactivity'' 
exploits the fact that watermark signals leave detectable traces when 
watermarked text is used by another model, enabling active training and 
context membership inference~\citep{sander2025detecting, jovanovic2025ward}.
However, to our knowledge, there is no thorough evaluation of how post-hoc watermarking through LLM paraphrasing performs across different data domains, such as prose versus verifiable text like code.
Moreover, unlike generation-time watermarking, where watermark embedding should not delay the generation, post-hoc watermarking allows trading additional computation for a better quality-detectability trade-off.
This opens up several axes to explore: model size, watermarking method, decoding strategy (and even generating multiple candidates), as well as detection strategy.

\begin{figure}[t!]
    \centering
    \includegraphics[width=0.8\linewidth, clip, trim={0 0.08cm 0 1.2cm}]{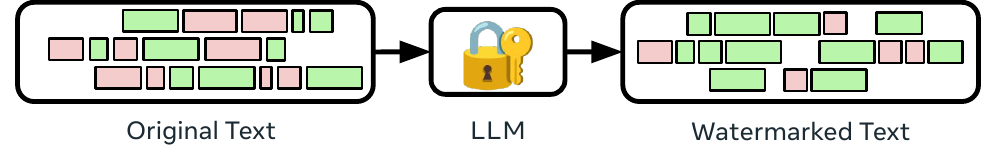}
    \caption{
        \textbf{Post-hoc text watermarking through watermarked LLM rephrasing. }
        We do empirical evaluations and analyze detection power, semantic fidelity, and correctness according to different design choices such as watermark scheme and available compute (through the paraphrasing model and the decoding strategy).
    }
    \label{fig:fig1}
\end{figure}

To address this gap, in this paper, we provide a comprehensive evaluation of post-hoc watermarking.
We find that current state-of-the-art LLM watermarking schemes applied in a post-hoc setting perform well on open-ended text, such as Wikipedia articles and books, achieving high detection power while preserving fidelity.
However, these methods are less effective on verifiable text like code, where the requirement to preserve correctness severely constrains paraphrasing freedom.
Regarding design choices, we find that larger models better preserve semantics, while smaller models are necessary to effectively hide strong watermarks. 
Furthermore, the simplest Gumbel-max approach~\citep{aaronson2023watermarking} dominates the Pareto frontier when classical random sampling is used.
For other watermarking methods, we find that beam search decoding significantly improves the quality-detectability trade-off.

In short, our main contributions are:
\begin{itemize}[leftmargin=*]
    \item \emph{A comprehensive evaluation:} We conduct the first large-scale study of post-hoc watermarking through LLM rephrasing (\autoref{fig:fig1}) across diverse domains, demonstrating that while current methods are effective for open-ended text, they struggle with verifiable formats like code.
    \item \emph{An analysis of design strategies:} We isolate the impact of model size, decoding, and watermarking schemes.
    For instance, we identify Gumbel-max as the robust Pareto-optimal choice (\autoref{fig:pareto_main}).
    \item \emph{An open-source research framework:} We release an easily modifiable codebase to facilitate research for post-hoc watermarking techniques.
\end{itemize}

\section{Related work}\label{sec:related}

\subsection{Post-Hoc Text Watermarking}\label{subsec:related_posthoc}

Early text watermarking altered text characteristics like characters or spacing~\citep{brassil1995electronic}. 
Other methods modify grammatical or syntactical structures via pre-established rules~\citep{topkara2005natural}, including synonym substitution~\citep{topkara2006hiding} and word reordering through passivization or topicalization~\citep{topkara2006words, topkara2006natural, meral2009natural}. 
Text steganography follows similar principles~\citep{winstein1998lexical, chapman2001practical, bolshakov2004method, shirali2008new, chang2014practical, xiang2017novel}. 
These edit-based systems typically exhibit low robustness and payload, \eg 1-2 bits per sentence~\citep{wilson2016avoiding}.
Deep learning methods have since been used for this task, for instance, with masked language models for steganography~\citep{ueoka2021frustratingly}, infilling models~\citep{yoo2023robust}, neural lexical substitution~\citep{qiang2023natural}, or encoder-decoder architectures~\citep{abdelnabi2021adversarial, zhang2024remark, xu2024robust}.

\subsection{Generation-Time Watermarking With Large Language Models}\label{subsec:mgenwatermark}

The first watermarks for machine-generated text date back to a method presumably used in Google Translate to filter automated translations from future training data~\citep{venugopal2011watermarking}.
For LLM-generated text, two concurrent approaches appeared shortly after the release of ChatGPT.
\cite{kirchenbauer2023watermark} bias a subset of the vocabulary, while \cite{aaronson2023watermarking} alter the sampling via the Gumbel trick. 
Both use pseudorandom seeds generated from a secret key and preceding tokens, enabling lightweight detection through statistical tests without access to the model.

Subsequent work explores improved tests and multi-bit watermarking~\citep{fernandez2023three, yoo2023advancing, yoo2024advancing, qu2024provably}, position-dependent seeds~\citep{christ2023undetectable, kuditipudi2023robust}, low-entropy optimizations~\citep{lee2023wrote, christ2023undetectable, huang2023optimal}, and semantic watermarks for better robustness~\citep{liu2023semantic, liu2024adaptive, fu2024watermarking, hou2023semstamp, hou2024k}. 
DiPMark~\citep{wu2023dipmark} provides distortion-free Green-Red watermarks, and MorphMark~\citep{wang2025morphmark} adaptively adjusts watermark strength based on the green token probability mass.
WaterMax~\citep{giboulot2024watermax} generates several chunks of tokens from the original LLM distribution and selects outputs with high watermark scores, which ensures that the original LLM distribution is preserved.
SynthID-Text~\citep{dathathri2024scalable} deploys tournament-based sampling in Google Gemini. 
\looseness=-1 Toolkits have also been introduced to benchmark these methods~\citep{piet2023mark, pan2024markllm}. 

We provide a comprehensive description of the schemes evaluated in this work in \autoref{app:method_overview}.

\subsection{Post-Hoc LLM Watermarks for Data Protection}
Recent works apply LLM watermarks to training or evaluation data via paraphrasing, similar to what we study in this work.
They exploit watermark radioactivity~\citep{sander2024watermarking}, \ie the detectable traces left when watermarked text is used for training. 
Applications include detection of texts used in retrieval-augmented generation (RAG)~\citep{jovanovic2025ward}, benchmark contamination detection~\citep{sander2025detecting}, and training data copyright~\citep{zhang2025leave}. 
Only Waterfall~\citep{lau2024waterfall} evaluates post-hoc watermarking through LLM paraphrasing, focusing on code (MBPP) and natural text (C4 and arXiv) for provenance detection.

These works demonstrate the utility of such methods for data protection, but do not evaluate it as a general watermarking method, nor do they characterize failure modes across text types and settings.

\section{Method}\label{sec:method}

\paragraph{\textbf{Post-hoc watermarking via paraphrasing.}}
The key idea is to paraphrase the input text using an LLM while applying watermarking during generation, as depicted in~\autoref{fig:fig1}.
The pipeline is as follows:
\begin{enumerate}
    \item Split input text into chunks (sentences or paragraphs).
    \item For each chunk, prompt an LLM to produce a paraphrase, given specific instructions (e.g., preserve named entities, minimal lexical change) and some previous context.
    \item During decoding, sample the next token with watermarking: for instance, favor tokens in a ``greenlist'' according to a strength parameter or any of the methods described in~\autoref{subsec:mgenwatermark}.
    \item Aggregate watermark scores across chunks and compute the detector statistic.
\end{enumerate}

\subsection{LLM Watermarking}

\paragraph{\textbf{Next token generation.}}
At each decoding step, the LLM computes logits $\bm{\ell} \in \mathbb{R}^{|\V|}$ over the vocabulary $\V$ conditioned on the preceding context.
These logits are converted into a probability distribution $\mathbf{p} = \text{softmax}(\bm{\ell}/T)$ via temperature scaling by $T$.

\paragraph{\textbf{Watermarked sampling.}}
Watermarking modifies the decoding process rather than sampling directly from $\mathbf{p}$.
This is done depending on a window $\mathbf{w}$ of the $k$ preceding tokens and a secret key $\sk$.
A pseudorandom function $\textsc{PRF}(\cdot)$ combines $\mathbf{w}$, $\sk$, and a candidate token $v$ to produce a value used for biasing the sampling of token $v$.
For instance, in the green-red method~\citep{kirchenbauer2023watermark}, we compute $\textsc{PRF}(\mathbf{w}, \sk, v)$ for all tokens $v \in \V$ to partition the vocabulary into a ``greenlist'' and a ``redlist,'' then bias the logits by adding $\delta$ to green tokens.
Other schemes alter the sampling mechanism directly (e.g., with the Gumbel-max trick~\citep{aaronson2023watermarking}).
This embeds a statistical signal into the generated text that is imperceptible to readers but detectable by the detector.
We provide a comprehensive description of the watermarking schemes evaluated in this work in \autoref{app:method_overview}.

\paragraph{\textbf{Detection and statistical test.}}
Detection recomputes the pseudorandom function for each group of tokens and aggregates the results into a test statistic.
Under the null hypothesis $\H_0$ ``not watermarked with this specific scheme'', the statistic should follow a known distribution, yielding a $p$-value for the probability of observing such a score or a more extreme score by chance.
Under the alternative hypothesis $\H_1$ ``watermarked text with this specific scheme'', the statistic should deviate significantly enough and lead to low $p$-values.
We flag text as watermarked if the $p$-value is below a detection threshold $\alpha$, which controls the false positive rate (FPR).

For instance, if $K$ (the test statistic here) is the observed number of green tokens out of $N$ scored tokens and this statistic follows under $\H_0$ a Binomial distribution with parameters $N$ and $\gamma$ (expected green ratio, typically $0.5$), then $p\text{-value} = \prob \left(X \geq K \mid X \sim \text{Binomial}(N, \gamma)\right)$.
As an example, if we observe $K=65$ green tokens out of $N=100$ with $\gamma=0.5$, the $p$-value is approximately $10^{-3}$; to flag this text as watermarked, we would need a FPR of $\alpha \geq 10^{-3}$.

An important practical consideration is \textit{deduplication}: because the pseudorandom function depends on the preceding $k$ tokens, repeated n-grams generate identical hashes.
This violates the statistical independence of the scores used when assuming the distribution of the test statistic under $\H_0$.
Aggregating scores only over unique watermark windows within the text~\citep{kirchenbauer2023watermark, fernandez2023three} is a good way to mitigate this issue and to ensure valid $p$-values.
But deduplication alone may not guarantee valid statistical tests because natural language inherently favors certain n-grams over others.
This bias can cause $p$-values to be artificially low or high even for unwatermarked text.
To address this, we ensure that under $\H_0$, $p$-values are approximately uniform on $U(0,1)$.
In practice, we test many candidate secret keys and select one for which the empirical $p$-value distribution on unwatermarked text is close to uniform.
We provide more details in~\autoref{app:hash}.

\subsection{Compute-Driven Flexibility}\label{subsec:compute_flexibility}

Operating in a post-hoc setting offers more flexibility than in-generation watermarks, which must comply with live-serving constraints such as latency, memory, and decoding speed~\citep{dathathri2024scalable}.
Here, we can use large or small models, run higher temperatures or beam search instead of simple sampling, and even employ multi-candidate selection like WaterMax~\citep{giboulot2024watermax}, as a function of available compute and quality targets.
We detail the specific higher-compute decoding (and detecting) methods we test below.

\paragraph{\textbf{Beam search for watermarking.}}
We use a beam search approach to improve the quality-detectability trade-off. 
Specifically, we maintain $B$ candidate sequences and expand each beam with $V$ candidates sampled from the watermarked probability distribution $\mathbf{p}_{\text{wm}}(\cdot \mid x_{<t})$, which is derived by applying the watermarking scheme (e.g., Maryland, SynthID, DIPMark, MorphMark) to the base model's logits. To select the top-$B$ beams at each step, we score candidates using log-probability under a reference distribution. We explore two scoring variants: \emph{unbiased scoring} uses the original model probabilities $\mathbf{p}_{\text{orig}}(\cdot \mid x_{<t})$ to favor sequences with lower perplexity relative to the base model, prioritizing text quality; \emph{biased scoring} uses the watermarked probabilities $\mathbf{p}_{\text{wm}}(\cdot \mid x_{<t})$ to favor sequences that are most probable under the watermarked distribution, potentially yielding stronger watermark signals.
We note that the latter was also explored for generation-time watermarking in~\citep{kirchenbauer2023reliability}.
For candidate generation, we consider both deterministic beam search (selecting top-$V$ tokens by $\mathbf{p}_{\text{wm}}$) and stochastic beam search (sampling $V$ tokens from $\mathbf{p}_{\text{wm}}$).

\paragraph{\textbf{Multi-candidate selection with WaterMax~\citep{giboulot2024watermax}}.}
Rephrasing is done chunk by chunk, with chunks of $L$ tokens.
For each chunk, we generate $m$ candidates $\{\tilde{y}^{(1)}, \dots, \tilde{y}^{(m)}\}$ \textit{without} applying any logit bias.
We then select the candidate that naturally maximizes the watermark score:
$y^* = \arg\max_{y} \text{Score}_{\text{wm}}(y)$.
This method is ``distortion-free'' regarding the sampling distribution but is typically infeasible in standard API usage due to cost (generating $L$ times more tokens per chunk).

\paragraph{\textbf{Entropy-aware detection.}}
At detection time, we compute the entropy $H_t$ of the model's predicted distribution at each token position $t$ in the watermarked text $H_t = - \sum_{v \in \V} p_t(v) \log p_t(v)$ where $p_t(v)$ is the model probability for token $v$ given the prefix up to $t-1$ in the watermarked text. 
We then apply an entropy filter: a token is included in the watermark score only if $H_t$ exceeds a chosen threshold $\tau$, similar to what is done in~\citet{lee2023wrote} for code. 
This focuses detection on high-entropy positions, where watermarking is more effective, and ignores low-entropy tokens. 
All entropy computations are performed on the watermarked text, as the original text is not available at detection time.
As a result, these entropy values differ from those at generation-time, when the model was conditioned.

\section{Experiments}\label{sec:results}

We conduct a comprehensive evaluation of post-hoc watermarking to assess: (1) \emph{detection power} (the ability to reliably identify watermarked text, measured through $p$-values) and (2) \emph{fidelity} (the preservation of meaning and quality through paraphrasing).
In~\autoref{subsec:exp_set_up}, we detail the experimental setup, including datasets, models, watermarking schemes, decoding strategies, and evaluation metrics.
In~\autoref{subsec:exp_pareto}, we compare watermarking methods on the quality-detectability Pareto frontier, finding that Gumbel-max dominates under standard sampling.
In~\autoref{subsec:model_scale}, we show that larger models preserve semantics better, but small/mid-size models offer sweet spots, with stronger watermark signals due to higher entropy.
In~\autoref{subsec:beam_search}, we demonstrate that beam search improves the Pareto frontier for all applicable methods.
In~\autoref{subsec:entropy_ablation}, we evaluate entropy-aware detection and find only modest gains.
In~\autoref{sec:code_exp_details}, we investigate post-hoc watermarking on code, revealing that correctness constraints limit detectability.
In~\autoref{subsec:multilingual}, we evaluate cross-lingual robustness and, in~\autoref{subsec:chunking}, the impact of chunking on long documents.

\subsection{Experimental Setup}\label{subsec:exp_set_up}

\paragraph{\textbf{Datasets.}}
We evaluate on three diverse corpora to capture different text characteristics and use cases:
\begin{itemize}
    \item \emph{Books:} Passages in English from the Gutenberg dataset~\citep{projectgutenberg}, containing books.
    These represent longer-form, literary texts with complex sentence structures.
    \item \emph{Wikipedia:} The beginning of randomly sampled Wikipedia articles. 
    These represent factual, encyclopedic texts with a formal writing style and different languages.
    \item \emph{Code:} Programming problems from the HumanEval~\citep{chen2021evaluating} and MBPP~\citep{austin2021program} benchmarks, consisting of Python functions with accompanying unit tests.
\end{itemize}

\paragraph{\textbf{Paraphrasing models and prompts.}}
We experiment with several dense open-weight instruct LMs as paraphrasers, spanning different families and sizes to study the impact of model scale on paraphrase quality and watermark preservation. 
Specifically, we use the instruct version of \textbf{LLaMA-3}~\citep{dubey2024llama} (1B, 3B, 8B, 70B), \textbf{Gemma-3}~\citep{gemma32025} (1B, 4B, 12B, 27B), \textbf{Qwen-2.5}~\citep{qwen25} (0.5B, 1.5B, 3B, 7B, 14B, 32B), and \textbf{SmolLM-2}~\citep{allal2025smollm2} (135M, 360M, 1.7B). 

To ensure the model outputs the watermarked text directly and without any comments, we use a structured system prompt and prefill the assistant's response with a prefix (e.g., ``Rephrased text:''). 
An example prompt used is:
\begin{center}
\small \textit{``You are a text rephrasing assistant. You must rephrase the given text while strictly preserving its original meaning, style, and structure. You must output only the rephrased text, with no explanations or commentary.''}
\end{center}

\paragraph{\textbf{Chunking strategy.}}
Large documents may be difficult to rephrase in a single pass, or may even exceed the model’s context window, so we split texts into chunks.
We compare two modes: \emph{full-context}, which processes the entire document at once, and \emph{context-aware chunking}, which processes each chunk separately while prepending one or two previously rephrased chunks as context to preserve coherence.

\paragraph{\textbf{Watermarking schemes.}}
We evaluate multiple watermarking methods (detailed in~\autoref{app:method_overview}), all with window size $2$, varying their hyperparameters to explore their detection-quality trade-offs.

\begin{itemize}
    \item \emph{Green-Red List~\citep{kirchenbauer2023reliability}:} We vary the bias $\delta \in \{1.0, 2.0, 4.0\}$.
    \item \emph{DiPMark~\citep{wu2023dipmark}:} We vary the reweighting parameter $\alpha \in \{0.2, 0.3, 0.4\}$.
    \item \emph{MorphMark~\citep{wang2025morphmark}:} We fix $\kappa = 10$ and vary $p_0 \in \{0, 0.05, 0.1, 0.2\}$.
    \item \emph{SynthID-Text~\citep{dathathri2024scalable}:} We vary the number of tournaments $k \in \{10, 20, 30\}$.
    We also compute the $p$-values from weighted and non-weighted scores as described in~\autoref{app:method_overview}.
    \item \emph{Gumbel-max~\citep{aaronson2023watermarking}:} Fixed by temperature and top-$p$.    
\end{itemize}

\paragraph{\textbf{Choice of private key.}}
We address the private key issue described in~\autoref{sec:method} by testing 50 candidate keys. 
For each combination of tokenizer, watermarking method (and depth for SynthID), we evaluate the keys on non-watermarked texts to estimate the $p$-value distribution under $H_0$.
We then select the key that minimizes the deviation from $U(0,1)$, as measured by the Kolmogorov–Smirnov statistic.
Next, we verify that the theoretical FPRs for these keys match the empirical FPRs on 1.5M Wikipedia documents (they contain up to 100k tokens, with an average of 715 tokens).
This ensures that the subsequent results under $H_1$ reflect the intrinsic performance of each scheme rather than artifacts of a particular key choice.
Further details are provided in~\autoref{tab:h0_key_sensitivity}.

\paragraph{\textbf{Entropy aware detection.}} For all methods, we evaluate entropy-aware detection by computing per-token entropy with the same model as used for rephrasing.
At detection, we only score tokens with entropy $\geq \tau$, with $\tau \in \{0, 0.2,0.4,0.6,\dots,1.8,2.0\}$.
We note that we only pass the watermarked text to compute the per-token entropy without the original text as context.

\paragraph{\textbf{Decoding strategies.}}
We explore different decoding approaches:
\begin{itemize}
    \item \emph{Nucleus sampling~\citep{holtzman2019curious}.} We vary temperature $T \in \{0.7, 1.0, 1.2\}$ and top-$p \in \{0.9, 0.95, 0.99\}$. 
    Higher temperature or top-$p$ should increase entropy at the cost of faithfulness.
    \item \emph{Beam search~\citep{sutskever2014sequence},} as detailed in~\autoref{sec:method}.
    We test beam widths $B \in \{3, 5, 10\}$ with the same number of candidates per beam.
    We experiment with selection criteria based on either the non-watermarked likelihood or the watermarked model's likelihood, and use stochastic (temperature $1$ and top-$p$ 0.95) or deterministic selection within beams.
    \item \emph{WaterMax~\citep{giboulot2024watermax}}, as detailed in~\autoref{sec:method}.
    We generate $L \in \{4, 8, 16\}$ non-watermarked candidate tokens per draft $m \in \{4, 8\}$ and select the one with the highest number of green tokens, and repeat. 
\end{itemize}

\paragraph{\textbf{Evaluation metrics.}}
We employ the following metrics to evaluate performance:
\begin{itemize}
    \item \emph{Detection:} We compute the $p$-value of the corresponding detection test, which can be read as the probability that a watermark score at least as high would happen if the text were not watermarked.
    \item \emph{Quality:} We evaluate rephrasing quality along three axes.
    \emph{Semantic similarity:} We use the BERTScore~\citep{zhang2020bertscore} to measure meaning preservation between original and rephrased texts.
    \emph{Naturalness:} We compute the average cross entropy of the rephrased text using Mistral-7B-v0.3~\citep{jiang2023mistral}, conditioned on the prefix ``\textit{This is a rephrased version of [ORIGINAL TEXT]:}'', capturing if the output is fluent and plausible.
    \emph{Length ratio:} We report the character-level length ratio between rephrased and original text, which should remain close to $1$.
    \item \emph{Functional Correctness (Code):} For code datasets, we measure the percentage of paraphrased functions that pass their unit tests (Pass@1).
\end{itemize}

In Sections~\ref{subsec:exp_pareto}, \ref{subsec:model_scale}, \ref{subsec:entropy_ablation}, \ref{subsec:beam_search} and~\ref{subsec:chunking}, each reported value corresponds to the median over 100 passages of approximately 800 characters ($\sim$200 tokens) sampled from Charles Dickens' novels in the Project Gutenberg dataset~\citep{projectgutenberg}.
We keep only rephrasings where the output length remains comparable to the original (length ratio $\in [0.75, 1.25]$), and report results only for settings where at least 70\% of passages satisfy this constraint.

As detailed in~\autoref{app:additional_results}, output length distributions are remarkably consistent across watermarking schemes and are primarily determined by the base model rather than the watermarking method.
While output length is arguably a quality signal in itself, we invite the reader to examine these distributions in the appendix, allowing the main analysis to focus on semantic fidelity and detection power within valid generation bounds.

\begin{table}[t!]
    \centering
    \caption{\textbf{Qualitative comparison on literary text.} 
    A sample from \textit{A Christmas Carol} by Charles Dickens, rephrased using Llama-3.2-3B-Instruct with the Gumbel-max scheme, $T=0.7$ and top-p = 0.95. 
    The output maintains the original tone and structure while achieving high semantic similarity and good watermark detectability.}
    \vspace{-0.2cm}
    \small
    \begin{tabularx}{\linewidth}{@{}YY@{}}
        \toprule
        \textbf{Original Text} & \textbf{Watermarked Rephrasing} \\
        \midrule
``You don't believe in me,'' observed the Ghost.\tnl

``I don't,'' said Scrooge.\tnl

``What evidence would you have of my reality beyond that of your senses?''\tnl

``I don't know,'' said Scrooge.\tnl

``Why do you doubt your senses?''\tnl

``Because,'' said Scrooge, ``a little thing affects them. A slight disorder of the stomach makes them cheats. You may be an undigested bit of beef, a blot of mustard, a crumb of cheese, a fragment of an underdone potato. There's more of gravy than of grave about you, whatever you are!''\tnl

Scrooge was not much in the habit of cracking jokes, nor did he feel, in his heart, by any means waggish then. The truth is, that he tried to be smart, as a means of distracting his own attention, and keeping down his terror; for the spectre's voice disturbed the very marrow in his bones.
To sit, staring at those fixed glazed eyes, in silence for a moment, would play, Scrooge felt, the very deuce with him. 
&
``You're not convinced of my existence,'' said the Ghost.\tnl

``I am not,'' replied Ebenezer Scrooge.\tnl

``What evidence do you have to confirm my reality beyond the limits of your perceptions?''\tnl

``I don't know,'' Scrooge replied.\tnl

``Why do you question the reliability of your senses?''\tnl

``Because,'' Scrooge said, ``a minor aberration can significantly impact them. A slight irregularity in digestion can cause them to be misled. You may be an undigested morsel of meat, a splash of mustard, a fragment of a partially cooked potato. There's more of gristle than of flesh about you, regardless of what you are!''\tnl

Scrooge was not accustomed to indulging in humor, nor did he feel that he was particularly witty at any point. The truth was, that he tried to appear intelligent, as a means of diverting his own attention and suppressing his fear; as a way to distract himself from the unsettling presence of the spectre.
To sit, staring at those fixed, unblinking eyes, in silence for a moment, would play, Scrooge felt, a cruel and unforgiving reminder of the truth.
\\
        \midrule
        \multicolumn{2}{c}{\footnotesize \textbf{Stats:} SBERT: 0.904 \statsep Ratio: 1.07  \statsep Perplexity: 2.61 \statsep $p$-val: $1.7 \times 10^{-5}$} \\
        \bottomrule
    \end{tabularx}
    \vspace{-0.3 cm}
    \label{tab:main_example}
\end{table}

\subsection{Quality-Detection Trade-Off}\label{subsec:exp_pareto} 

\begin{figure}[b!]
    \centering
    \includegraphics[width=\linewidth]{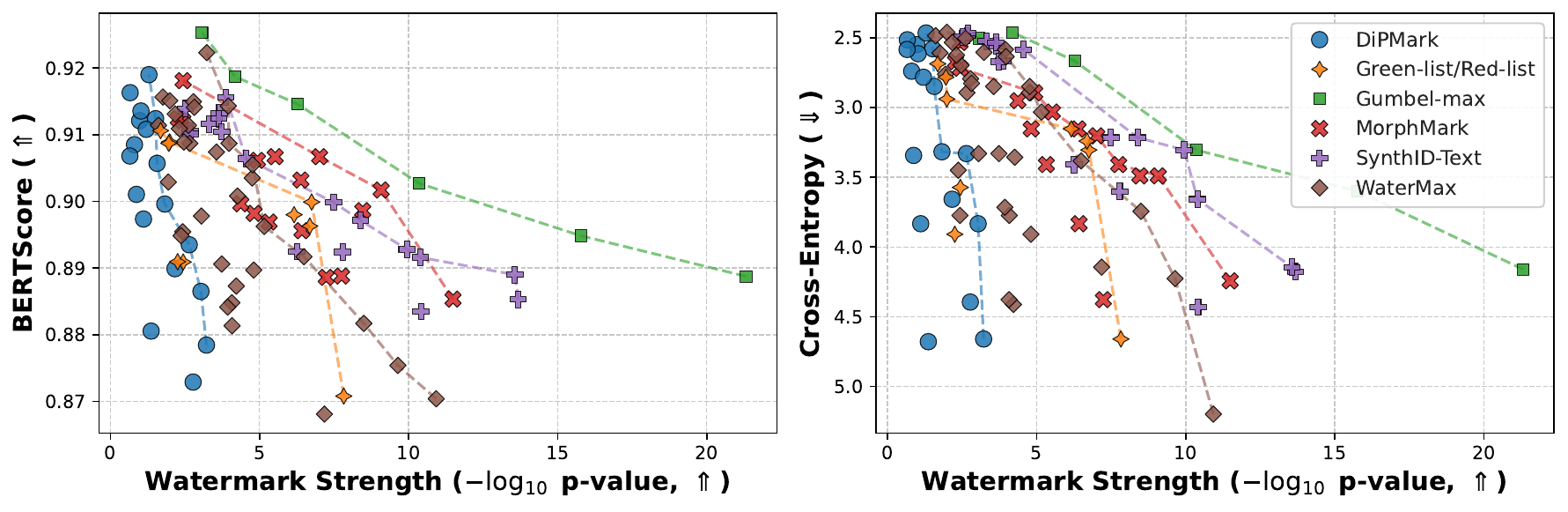}
    \vspace{-0.8cm}
    \caption{\textbf{Pareto fronts of watermarking methods} showing the trade-off between quality and watermark strength using Llama-3.2-3B-Instruct. Each point corresponds to a different parameter configuration, with values representing medians across 100 rephrased passages.
    Experimental details are given in Sections~\ref{subsec:exp_set_up} and~\ref{subsec:exp_pareto}.
    }
    \vspace{-0.2cm}
    \label{fig:pareto_main}
\end{figure}

We first evaluate the fundamental trade-off between rephrasing quality and 
detection strength in~\autoref{fig:pareto_main}, using Llama-3.2-3B-Instruct 
with random sampling.
Each point corresponds to a distinct parameter 
configuration as described in~\autoref{subsec:exp_set_up}.
Optimal methods occupy the top-right region of the plot, combining high text quality with strong detectability.
The Gumbel-max method appears to dominate this Pareto frontier, followed by MorphMark and SynthID. 

We also present a qualitative example in~\autoref{tab:main_example}, where we rephrase text using Llama-3.2-3B-Instruct with the Gumbel-max watermark with $T=0.7$ and $p=0.95$. 
The rephrasing maintains the original tone, structure, and factual content. 
Additional examples are provided in~\autoref{tab:appendix_examples} of~\autoref{app:qualitative}.

\subsection{Impact of Model Family and Size}\label{subsec:model_scale}

\begin{figure}[b!]
    \centering
    \includegraphics[width=\linewidth]{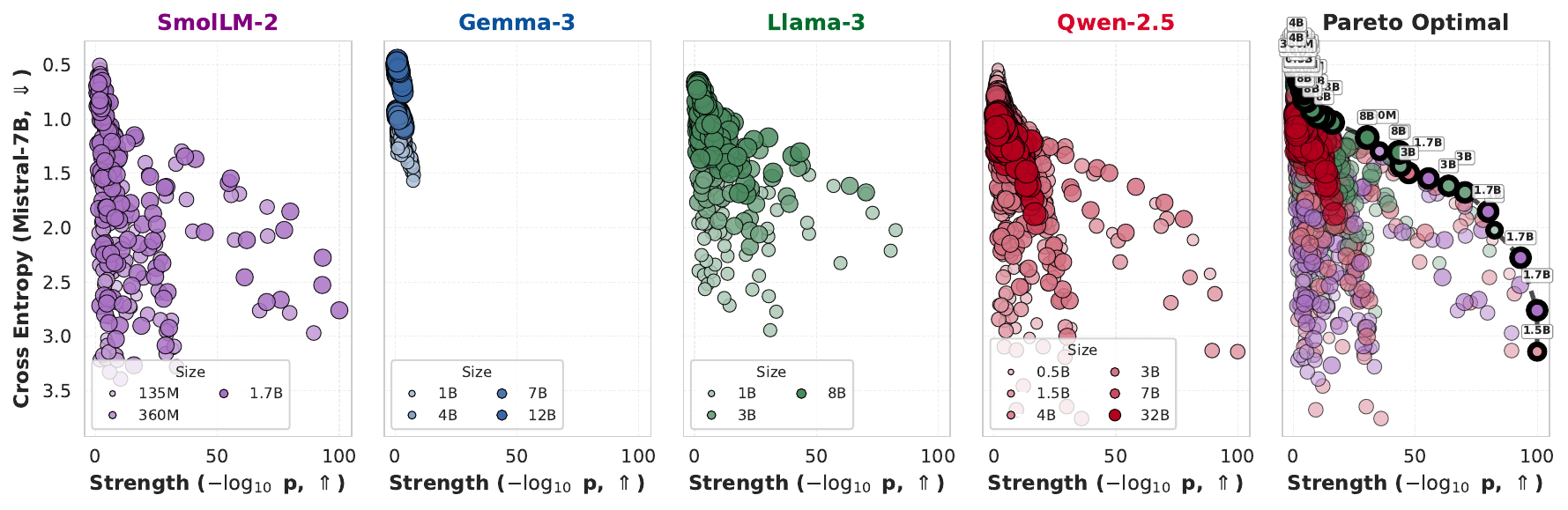}
    \caption{\textbf{Impact of model family and size.} Cross Entropy  vs.\ watermark strength.
    Larger models improve quality for a given watermark strength, but small models are necessary to reach high strengths. 
    All families are comparable, except Gemma-3 which is not suitable.
    Experimental details are given in Sections~\ref{subsec:exp_set_up} and~\ref{subsec:model_scale}.
    }
    \label{fig:model_scale}
\end{figure}

We investigate whether larger models are better suited for post-hoc watermarking.
\autoref{fig:model_scale} shows the trade-off curves by model family (SmolLM, Gemma, Llama, Qwen) and size.
Regarding performance, we observe that larger models preserve semantics better: across all families, increasing model size (darker points) shifts the clusters upward.
For example, within the Llama family, Llama-3.1-8B (dark green) attains lower cross entropy than the 3B and 1B variants (lighter teal) at comparable watermark strengths.
However, larger models struggle to reach low $p$-values compared to smaller models: they populate the left rather than the right of the plots.
Gemma~3 models are essentially absent from the high-detectability region, as their outputs naturally exhibit very low entropy, even for the smallest sizes.
Overall, when strong watermarking is required, only small models appear on the frontier, whereas larger models dominate in the lower-strength regime.
We see in~\autoref{subsec:code} that the issue with large models persists even if the temperature is increased.

Note that these results are conditioned on the length filtering described in~\autoref{subsec:exp_set_up}; thus, the presence of a data point implies enough successful valid generation for the corresponding combination of model/scheme/parameters.
For instance, smaller models with stronger watermarks fail to meet length constraints more frequently.
We refer the reader to \autoref{app:additional_results} to assess the generation stability of each configuration, which cannot be inferred directly from the figure.

\subsection{Decoding Strategies: Beam Search vs. Sampling vs. WaterMax}\label{subsec:beam_search}

\begin{figure}[b!]
    \centering
    \includegraphics[width=\linewidth]{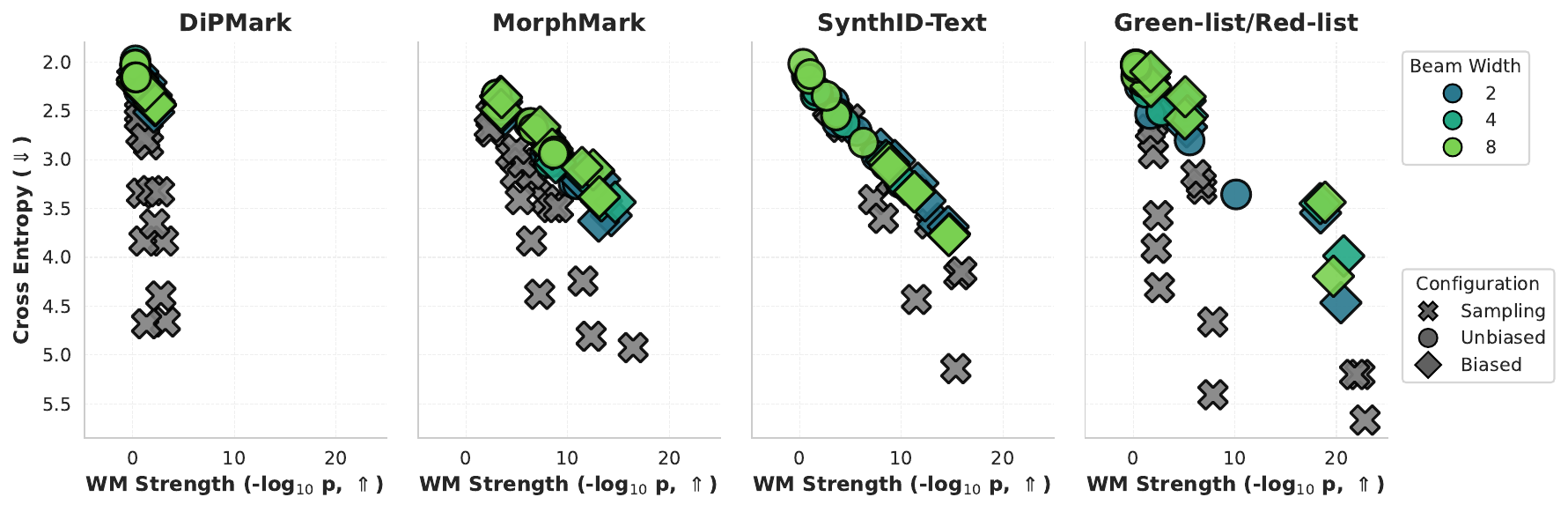}
    \vspace{-0.4cm}
    \caption{\textbf{Beam search improves the Pareto frontier.} Cross entropy vs.\ watermark strength for suitable methods.
    Beam search, especially with biased scoring (see~\autoref{sec:method}), shifts the frontier upward, substantially improving rephrasing quality at a fixed watermark strength.
    Experimental details are given in Sections~\ref{subsec:exp_set_up} and~\ref{subsec:beam_search}.
    }
    \label{fig:beam_search}
\end{figure}

We explore in~\autoref{fig:beam_search} whether beam search can find better watermarked sequences than random sampling.
We watermark with Llama-3.2-3B-Instruct, and decode with or without beam search, for each suitable watermarking method: Green-red, Synth-ID, MorphMark, and DiPMark (since Gumbel-max is deterministic, beam search cannot be applied with it).

We observe that beam search consistently improves the Pareto frontier, \emph{especially with biased scoring} (see~\autoref{sec:method}): it shifts the results upward, substantially improving rephrasing quality at a fixed watermark strength.
Notably, while Gumbel-max sampling appears to be the best option when used with random sampling in ~\autoref{fig:pareto_main}, other methods can be substantially improved through beam search.

WaterMax~\citep{giboulot2024watermax} results are shown in brown in~\autoref{fig:pareto_main}.
Surprisingly, we found that WaterMax achieved weak detectability across the hyperparameters we tested, making it difficult to achieve strong watermarking detection in the post-hoc setup.

\subsection{Entropy-Aware Detection}\label{subsec:entropy_ablation}

\begin{figure}[b!!]
    \centering
    \includegraphics[width=\linewidth]{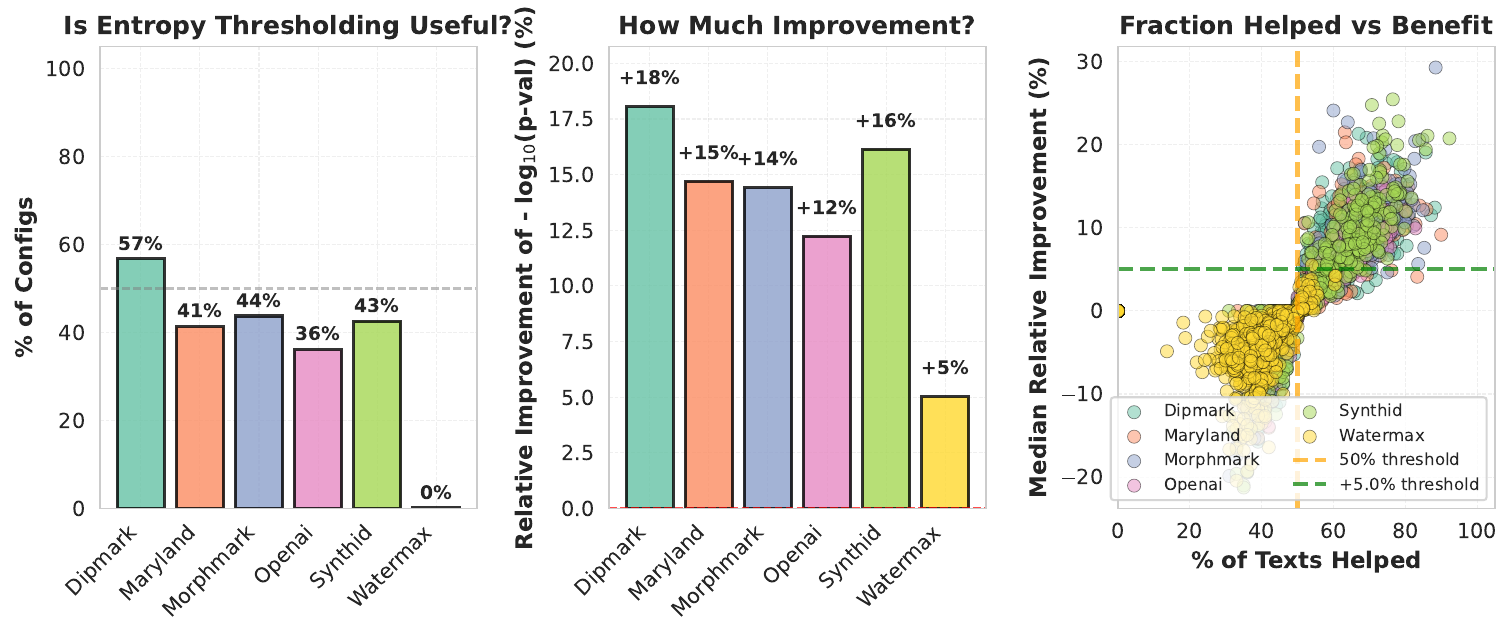}
    \caption{
       \textbf{Effect of entropy-aware detection.}
    \emph{Left:} Share of configurations for which some threshold improves detection by more than $5\%$ on at least half of the texts.
    \emph{Middle:} Median relative improvement for those configurations.
    \emph{Right:} For every configuration, fraction of texts helped at optimal threshold vs. median improvement; dashed lines mark the $50\%$ and $+5\%$ criteria.
    \looseness=-1
    Experimental details are given in Sections~\ref{subsec:exp_set_up} and~\ref{subsec:entropy_ablation}.
    }
    \label{fig:entropy_ablation}
\end{figure}

Instead of scoring every token, we now filter out low-entropy tokens at detection time, as motivated in~\autoref{sec:method} and detailed in~\autoref{subsec:exp_set_up}.
In~\autoref{fig:entropy_ablation}, we fix the rephrasing model to Llama-3.2-3B and, for each method and watermarking configuration, test whether there exists an entropy threshold that improves detectability by more than $5\%$ on at least 50 of the 100 texts of Charles Dickens.

The left plot reports, for each watermarking method, the proportion of configurations for which such a threshold exists.
We see that most methods have only $30$–$40$ successful configurations.
The middle plot shows, for those successful configurations, the actual gain in detection performance, which never exceeds $20\%$.
The right panel presents the corresponding non-aggregated statistics.

We see that WaterMax is different: any entropy threshold degrades detectability.
This is because WaterMax optimizes the watermark score at the \emph{sentence} level rather than the token level; filtering by token entropy does not expose additional signal, it only reduces the number of scored tokens and thus the available statistical evidence.
One other and related property of WaterMax is that it cannot be used for active dataset inference through radioactivity. 
This is explained in detail in ~\autoref{app:additional_results}.

\subsection{Post-hoc Code Watermarking}\label{subsec:code}

\paragraph{\textbf{Experimental details.}}\label{sec:code_exp_details}
We now conduct experiments on HumanEval and MBPP to evaluate the feasibility of post-hoc watermarking on structured and verifiable text, where watermarking requires preserving the syntax and the function of the code.

\emph{Data preparation and preprocessing.}
For each problem instance, we concatenate the provided prompt (\eg function signature and docstring) and the canonical solution. 
We make sure that the last top-level function definition of the code is the target for execution by moving it if necessary.
We show in App.~\ref{app:code-tasks} an example of a HumanEval task.
Overall, this dataset comprises 164 HumanEval and 974 MBPP code samples, with average lengths of 180 and 80 tokens, respectively. The code samples range from approximately 20 to 500 tokens in length.

\emph{Watermarking pipeline.}
The watermarking process is applied only to the code (we exclude unit tests to preserve the evaluation). 
Since watermarking may inadvertently rename function identifiers, we implement a post-processing step that parses the output and restores the original entry point name required by the benchmark test harness.
At test time, we concatenate the watermarked code with the test harness to evaluate functional correctness.

\paragraph{Metrics for code watermarking.}
First, we measure \emph{functional correctness}: the fraction of watermarked code that passes the tests (pass@1). 
If pass@1 is low, one could regenerate with different random seeds until obtaining functional code, but this assumes tests are available and it is cumbersome; ideally, pass@1 should be high.
Second, we measure the \emph{detection power}: the true positive rate (TPR) at a fixed false positive rate (FPR), and/or the distribution of $\log_{10}$ $p$-values. 
These can be computed among all rephrased code, or more interestingly, among only the ones that passed the tests.
In practice, the latter is a better indicator of the watermark power, since codes that do not pass tests often exhibit degenerative patterns (repeated code or text at the end) which artificially increase the TPR.

\begin{table}[t!]
    \centering
    \caption{
        \textbf{Qualitative example of code watermarking} on a sample from MBPP (using Llama-3.1-8B at temperature 1.4 with Gumbel-max watermarking, top-$p=0.95$).
        The post-hoc rephrasing here refactors variable names to ensure detection (low p-value) while maintaining correctness.
        }
    \small
    \begin{tabularx}{\linewidth}{@{}YY@{}}
        \toprule
        \textbf{Original Text} & \textbf{Watermarked Rephrasing} \\
        \midrule
        \ttfamily\scriptsize
        \# A python function to identify non-prime numbers. \newline
        import math\ \newline
        def is\_not\_prime(n):\ \newline
        \hspace*{1em}result = False\ \newline
        \hspace*{1em}for i in range(2,int(math.sqrt(n)) + 1):\ \newline
        \hspace*{2em}if n \% i == 0:\ \newline
        \hspace*{3em}result = True\ \newline
        \hspace*{1em}return result \newline
        
        & 
        \ttfamily\scriptsize
        \# A python function to identify non-prime numbers\ \newline
        import math\ \newline
        \ \newline
        def is\_not\_prime(number):\ \newline
        \hspace*{1em}verdict = False\ \newline
        \hspace*{1em}for divisor in range(2, int(math.sqrt(number)) + 1):\ \newline
        \hspace*{2em}if number \% divisor == 0:\ \newline
        \hspace*{3em}verdict = True\ \newline
        \hspace*{1em}return verdict\ \\
        \midrule
            \footnotesize $p_{\text{value}}$: $0.12$ \statsep Tokens: $48$ \statsep Correct: \cmark & 
            \footnotesize  $p_{\text{value}}$: $3.29 \times 10^{-3}$ \statsep Tokens: $57$ \statsep Correct: \cmark
        \\
        \bottomrule
    \end{tabularx}
    \label{tab:code_example}
\end{table}

\paragraph{\textbf{Comparing watermarking methods.}}

\begin{figure}[b!!]
    \centering
    \includegraphics[width=\linewidth]{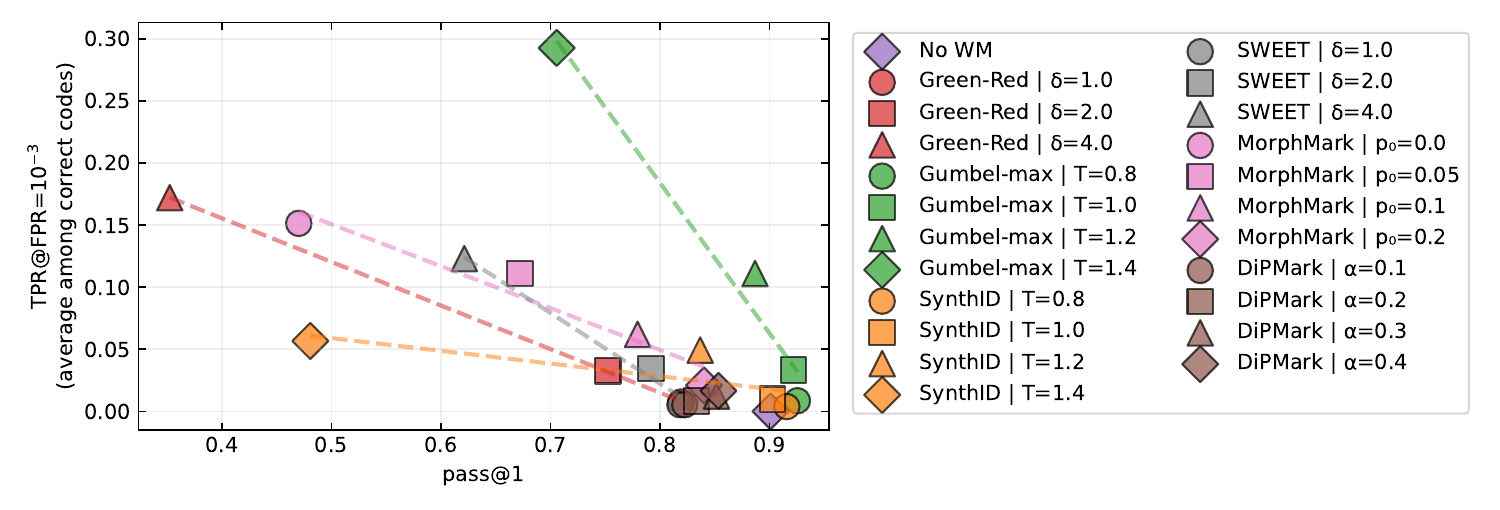}
    \caption{
        \textbf{Comparison of methods on code} using Llama-3.1-8B.
        We report pass@1 (functional correctness) vs.\ TPR at FPR=$10^{-3}$ among correct codes, averaged over HumanEval and MBPP.
        Different markers and colors correspond to different watermarking methods and their respective hyperparameters.
        Consistent with our findings in literary English, Gumbel-max watermarking achieves the best trade-off between utility and detectability.
    }
    \label{fig:code_pareto_8b}
\end{figure}

We compare the watermarking methods using Llama-3.1-8B.
\autoref{fig:code_pareto_8b} shows the trade-off between functional correctness (pass@1) and detection power (TPR at FPR=$10^{-3}$ among correct codes) for each method.
Gumbel-max watermarking achieves the best Pareto frontier, offering the most favorable trade-off between utility and detectability.
SynthID performs well in moderate temperature regimes but breaks down at higher temperatures, which limits its ability to achieve high TPR values.
MorphMark and DiPMark show competitive performance at intermediate operating points, while Green-Red and SWEET exhibit steeper degradation in pass@1 as watermark strength increases.
Overall, no method achieves both high pass@1 and high TPR simultaneously, confirming the inherent difficulty of watermarking code while preserving functionality.

\paragraph{\textbf{Comparing models.}}

\begin{figure}[b!]
    \centering
    \begin{subfigure}[b]{0.49\linewidth}
        \centering
        \includegraphics[width=\linewidth]{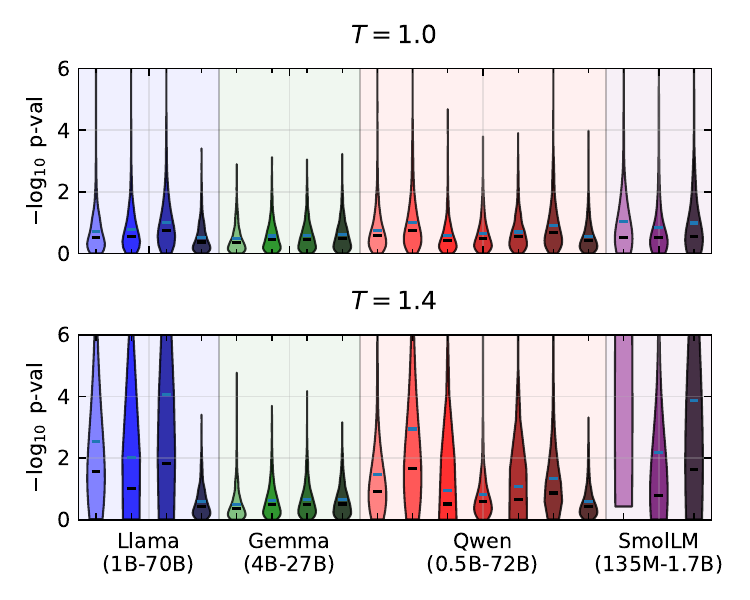}
        \caption{}
        \label{fig:code_models_violin}
    \end{subfigure}
    \hfill
    \begin{subfigure}[b]{0.49\linewidth}
        \centering
        \includegraphics[width=\linewidth]{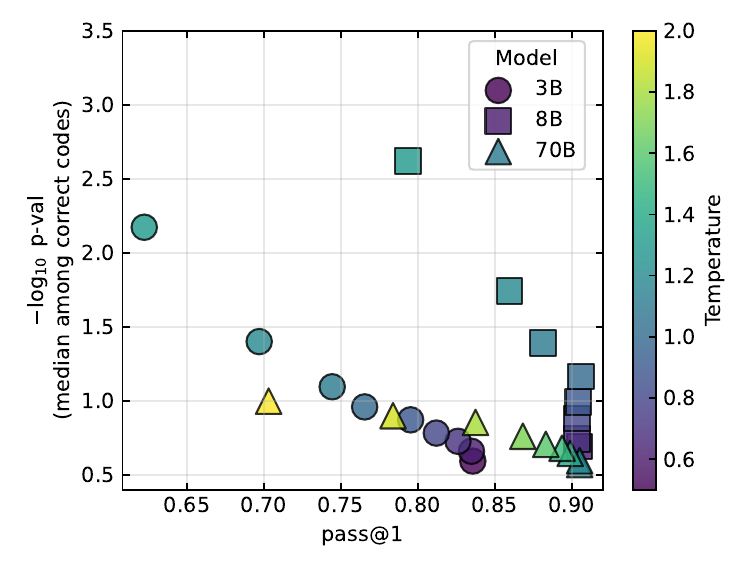}
        \caption{}
        \label{fig:code_pareto_models}
    \end{subfigure}
    \caption{
        \textbf{Evaluation of post-hoc watermarking across models} using Gumbel-max (top-$p$=0.95).
        (a)~Distribution of -$\log_{10}$ p-value for watermarked codes that pass the functional correctness test (bars indicate mean and median). Different colors represent model families, with darker shades indicating larger models.
        Corresponding values for Llama models are detailed in \autoref{tab:llama_watermark}.
        (b)~Pass@1 and detection power (median -$\log_{10}$ p-value among correct codes) for Llama 3B, 8B and 70B models across temperatures. 
        Different model families and sizes behave differently in terms of detection power at fixed temperatures, and interestingly, smaller models can achieve better Pareto optimality than larger models.
    }
    \label{fig:code_models}
\end{figure}

We compare model families and sizes using Gumbel-max watermarking (top-$p$=0.95).
Figure~\ref{fig:code_models_violin} shows that larger models lead to lower detection power (lower -$\log_{10}$ p-values), as their outputs are less entropic and leave less room for watermarking.
\autoref{tab:llama_watermark} confirms this: Llama-70B achieves median -$\log_{10}$ p-values around 0.5-0.7, while smaller models reach values near 2 at $T=1.4$.
However, higher detection power in smaller models comes at the cost of functional correctness; for example, at $T=1.4$, Llama-8B achieves a TPR of 0.29 among correct codes but a pass@1 of only 0.71.

To further investigate this trade-off, we sweep the temperature in~\autoref{fig:code_pareto_models}.
Interestingly, the smaller 8B model achieves a better Pareto frontier than the larger 70B model, attaining higher detection power at equivalent pass@1 levels.
This suggests that while larger models produce higher-quality code, their reduced entropy limits watermark effectiveness, making smaller models more suitable for post-hoc code watermarking when balancing functionality and detectability.

\begin{table}[t!]
\centering
\caption{
    \textbf{Post-hoc watermarking on code with Llama models} using Gumbel-max (top-$p$=0.95). We report pass@1, median $\log_{10}$ $p$-value ($\uparrow$), and TPR at FPR=$10^{-3}$ for different model sizes and temperatures ($T$). 
    Metrics are shown over all samples (``All'') and restricted to samples where the watermarked code passes functional tests (``Passed'').
    Increasing model size increases pass@1 but decreases detection power, with smaller models achieving better Pareto optimality.
}
\label{tab:llama_watermark}
\resizebox{\textwidth}{!}{%
\begin{tabular}{l ccccc ccccc ccccc}
\toprule
 & \multicolumn{5}{c}{$T = 1.0$} & \multicolumn{5}{c}{$T = 1.2$} & \multicolumn{5}{c}{$T = 1.4$} \\
\cmidrule(lr){2-6} \cmidrule(lr){7-11} \cmidrule(lr){12-16}
Model & pass@1 & \multicolumn{2}{c}{-$\log_{10}$ p-val} & \multicolumn{2}{c}{TPR@$10^{-3}$} & pass@1 & \multicolumn{2}{c}{-$\log_{10}$ p-val} & \multicolumn{2}{c}{TPR@$10^{-3}$} & pass@1 & \multicolumn{2}{c}{-$\log_{10}$ p-val} & \multicolumn{2}{c}{TPR@$10^{-3}$} \\
 &  & All & Passed & All & Passed &  & All & Passed & All & Passed &  & All & Passed & All & Passed \\
\midrule
Llama-3.2-1B-Instruct & 0.59 & 0.60 & 0.51 & 0.03 & 0.01 & 0.44 & 1.32 & 0.88 & 0.22 & 0.07 & 0.16 & 9.10 & 1.56 & 0.72 & 0.29 \\
Llama-3.2-3B-Instruct & 0.81 & 0.60 & 0.56 & 0.01 & 0.01 & 0.73 & 0.87 & 0.75 & 0.08 & 0.05 & 0.53 & 1.84 & 1.00 & 0.34 & 0.13 \\
Llama-3.1-8B-Instruct & 0.92 & 0.74 & 0.74 & 0.04 & 0.03 & 0.89 & 1.16 & 1.11 & 0.12 & 0.11 & 0.71 & 2.31 & 1.83 & 0.41 & 0.29 \\
Llama-3.3-70B-Instruct & 0.92 & 0.38 & 0.37 & 0.00 & 0.00 & 0.92 & 0.41 & 0.40 & 0.00 & 0.00 & 0.90 & 0.44 & 0.44 & 0.00 & 0.00 \\
\bottomrule
\end{tabular}}
\end{table}

\subsection{Multi-Lingual Robustness on Wikipedia}
\label{subsec:multilingual}

To evaluate the generalization of post-hoc watermarking beyond English, we apply the method to paragraphs of Wikipedia articles in multiple languages. 
In the following, we use Llama-3.1-8B-Instruct, with Gumbel-max watermarking with varying temperature settings (0.8, 1.0, 1.2) and fixed top-$p=0.95$.
\autoref{tab:multilingual_pareto} compares the performance across English (en), Spanish (es), French (fr), and Russian (ru).
To not conflate the effect of the number of tokens used for scoring with language, we only consider outputs that contain between 400 and 600 tokens, and aggregate results over 1k samples.

\begin{table}[t!]
\centering
\caption{
\textbf{Evaluation of post-hoc watermarking for different languages.} We report the median log$_{10}$ p-value and the SBERT score for semantic similarity, at varying temperatures of the generation. 
While not impossible, watermarking languages other than
English comes at a steeper cost on semantic quality.
}\label{tab:multilingual_pareto}
\small
\begin{tabularx}{\linewidth}{l *{8}{>{\centering\arraybackslash}X}}
\toprule
Temperature & \multicolumn{2}{c}{en} & \multicolumn{2}{c}{es} & \multicolumn{2}{c}{fr} & \multicolumn{2}{c}{ru} \\
\cmidrule(lr){2-3} \cmidrule(lr){4-5} \cmidrule(lr){6-7} \cmidrule(lr){8-9}
 & $-\log_{10} p$ & SBERT & $-\log_{10} p$ & SBERT & $-\log_{10} p$ & SBERT & $-\log_{10} p$ & SBERT \\
\midrule
0.80 & 1.832 & 0.960 & 1.608 & 0.949 & 0.810 & 0.946 & 1.355 & 0.908 \\
1.00 & 3.813 & 0.954 & 4.408 & 0.940 & 2.809 & 0.937 & 4.976 & 0.888 \\
1.20 & 8.147 & 0.946 & 20.098 & 0.921 & 19.705 & 0.911 & 23.689 & 0.836 \\
\bottomrule\end{tabularx}\end{table}

The results indicate a performance gap: while non-English languages (Spanish, French, Russian) can be watermarked, they suffer from a steeper trade-off. 
Achieving high detection strength in these languages requires a larger sacrifice in semantic quality compared to English. 
This suggests that the logits of the paraphrasing model are less easy to manipulate in languages for which the model was not primarily trained, since most training data is in English. 
It may also reflect the fact that our semantic similarity metric (SBERT) is more accurate for English than for other languages.

\subsection{Impact of Chunking Strategy on Long Documents}
\label{subsec:chunking}

To assess the impact of chunking as described in~\autoref{sec:method}, we compare full-context processing with context-aware chunking (500-token chunks, with up to 1000 context tokens from previously rephrased chunks) on documents of varying length (500-4000 tokens), using Dickens novel excerpts and Llama-3.2-3B-Instruct.

For each document length, results are averaged over 5 independent texts.
\autoref{tab:chunking_impact} shows that full-context processing increasingly summarizes the content, while chunking better preserves length and yields stronger detection at better semantic similarity.
\looseness=-1 This indicates that context-aware chunking is crucial for reliable watermarking of important document lengths.

\begin{SCtable}[1][b!]
    \small
    \begin{tabular}{lcccccc}
        \toprule
        \multirow{2}{*}{\textbf{Size}} & \multicolumn{2}{c}{\textbf{Length Ratio}} & \multicolumn{2}{c}{\textbf{Detection ($-\log_{10} p$)}} & \multicolumn{2}{c}{\textbf{Similarity}} \\
        \cmidrule(lr){2-3} \cmidrule(lr){4-5} \cmidrule(lr){6-7}
         & \textit{Full} & \textit{Chunked} & \textit{Full} & \textit{Chunked} & \textit{Full} & \textit{Chunked} \\
        \midrule
         500  & 0.91 & 0.91 & 19.3 & 19.3 & 0.94 & 0.94 \\
        1500  & 1.15 & 0.97 & 49.5 & 148.8 & 0.85 & 0.91 \\
        2500  & 0.78 & 0.98 & 56.8 & 195.5 & 0.91 & 0.91 \\
        4000  & 0.70 & 0.86 & 24.9 & 198.3 & 0.84 & 0.91 \\
        \bottomrule
    \end{tabular}
    \caption{\textbf{Context-aware chunking vs.\ full-context.}
    Comparison of full-context and context-aware chunked rephrasing (500-token chunks with up to 1000 context tokens) on documents of varying length, averaged over 5 excerpts per length.
    Experimental details are given in~\autoref{subsec:exp_set_up} and~\autoref{subsec:chunking}.
    }
    \label{tab:chunking_impact}
\end{SCtable}

\section{Conclusion}\label{sec:conclusion}

We presented a comprehensive evaluation of post-hoc text watermarking via LLM rephrasing, a paradigm that enables embedding traceable signals into existing text. 
Unlike generation-time constraints, the post-hoc setting allows the allocation of additional compute to optimize the trade-off between text quality and watermark detectability.

Our experiments yielded several findings. 
First, the simplest Gumbel-max scheme~\citep{aaronson2023watermarking} achieved better trade-offs than all other tested methods under random sampling. 
We note that in the generation-time watermarking literature, Gumbel-max is criticized for its deterministic nature: it fixes the randomness of generation, so identical prompts produce identical outputs, which can be problematic in production.
In our post-hoc setting, however, this limitation is less consequential.
We also observe that all other schemes benefit substantially from beam search.
Second, to achieve high watermark strength, smaller, more entropic models outperformed larger models run at high temperature.
Third, entropy filtering at detection time provided only marginal gains while introducing additional complexity, making it a less practical option in typical deployment scenarios.

A limitation of our study, and of the field at large, is the reliance on automated metrics such as perplexity and BERTScore to approximate text quality and semantic preservation.
Much like in image and audio watermarking, these proxies cannot fully capture the nuances of human perception, particularly across different languages where robust human evaluations remain essential.
To address this limitation, we extended our analysis to post-hoc code watermarking, where execution-based correctness provides an objective, ground-truth measure of utility.
Our results reveal that while watermarking is feasible in this domain, strict correctness constraints reduce the available capacity for watermark embedding compared to natural language.

\clearpage
\bibliographystyle{ieeenat_fullname}
\bibliography{references}

\begin{thebibliography}{63}
\providecommand{\natexlab}[1]{#1}
\providecommand{\url}[1]{\texttt{#1}}
\expandafter\ifx\csname urlstyle\endcsname\relax
  \providecommand{\doi}[1]{doi: #1}\else
  \providecommand{\doi}{doi: \begingroup \urlstyle{rm}\Url}\fi

\bibitem[Aaronson and Kirchner(2023)]{aaronson2023watermarking}
Scott Aaronson and Hendrik Kirchner.
\newblock Watermarking {GPT} outputs, 2023.

\bibitem[Abdelnabi and Fritz(2021)]{abdelnabi2021adversarial}
Sahar Abdelnabi and Mario Fritz.
\newblock Adversarial watermarking transformer: Towards tracing text provenance with data hiding.
\newblock In \emph{2021 IEEE Symposium on Security and Privacy (SP)}, pages 121--140. IEEE, 2021.

\bibitem[Allal et~al.(2025)Allal, Lozhkov, Bakouch, Blázquez, Tunstall, Piqueres, Marafioti, Almubarak, Mangrulkar, Belkada, and von Werra]{allal2025smollm2}
Loubna~Ben Allal, Anton Lozhkov, Elie Bakouch, Gabriel~Martín Blázquez, Lewis Tunstall, Agustín Piqueres, Andres Marafioti, Khalid Almubarak, Sourab Mangrulkar, Younes Belkada, and Leandro von Werra.
\newblock Smollm2: When smol goes big -- data-centric training of a small language model, 2025.

\bibitem[Austin et~al.(2021)Austin, Odena, Nye, Bosma, Michalewski, Dohan, Jiang, Cai, Terry, Le, and Sutton]{austin2021program}
Jacob Austin, Augustus Odena, Maxwell Nye, Maarten Bosma, Henryk Michalewski, David Dohan, Ellen Jiang, Carrie Cai, Michael Terry, Quoc Le, and Charles Sutton.
\newblock Program synthesis with large language models, 2021.

\bibitem[Bolshakov(2004)]{bolshakov2004method}
Igor~A Bolshakov.
\newblock A method of linguistic steganography based on collocationally-verified synonymy.
\newblock In \emph{International Workshop on Information Hiding}, pages 180--191. Springer, 2004.

\bibitem[Brassil et~al.(1995)Brassil, Low, Maxemchuk, and O'Gorman]{brassil1995electronic}
Jack~T Brassil, Steven Low, Nicholas~F Maxemchuk, and Lawrence O'Gorman.
\newblock Electronic marking and identification techniques to discourage document copying.
\newblock \emph{IEEE Journal on Selected Areas in Communications}, 13\penalty0 (8):\penalty0 1495--1504, 1995.

\bibitem[Chang and Clark(2014)]{chang2014practical}
Ching-Yun Chang and Stephen Clark.
\newblock Practical linguistic steganography using contextual synonym substitution and a novel vertex coding method.
\newblock \emph{Computational linguistics}, 40\penalty0 (2):\penalty0 403--448, 2014.

\bibitem[Chapman et~al.(2001)Chapman, Davida, and Rennhard]{chapman2001practical}
Mark Chapman, George~I Davida, and Marc Rennhard.
\newblock A practical and effective approach to large-scale automated linguistic steganography.
\newblock In \emph{International Conference on Information Security}, pages 156--165. Springer, 2001.

\bibitem[Chen et~al.(2021)]{chen2021evaluating}
Mark Chen et~al.
\newblock Evaluating large language models trained on code.
\newblock \emph{arXiv}, 2021.

\bibitem[Christ et~al.(2023)Christ, Gunn, and Zamir]{christ2023undetectable}
Miranda Christ, Sam Gunn, and Or Zamir.
\newblock Undetectable watermarks for language models.
\newblock \emph{Cryptology ePrint Archive}, 2023.

\bibitem[Crothers et~al.(2022)Crothers, Japkowicz, and Viktor]{crothers2022machine}
Evan Crothers, Nathalie Japkowicz, and Herna Viktor.
\newblock Machine generated text: A comprehensive survey of threat models and detection methods.
\newblock \emph{arXiv preprint arXiv:2210.07321}, 2022.

\bibitem[Dathathri et~al.(2024)Dathathri, See, Ghaisas, Huang, McAdam, Welbl, Bachani, Kaskasoli, Stanforth, Matejovicova, et~al.]{dathathri2024scalable}
Sumanth Dathathri, Abigail See, Sumedh Ghaisas, Po-Sen Huang, Rob McAdam, Johannes Welbl, Vandana Bachani, Alex Kaskasoli, Robert Stanforth, Tatiana Matejovicova, et~al.
\newblock Scalable watermarking for identifying large language model outputs.
\newblock \emph{Nature}, 634\penalty0 (8035):\penalty0 818--823, 2024.

\bibitem[DeepMind(2025)]{gemma32025}
Google DeepMind.
\newblock Gemma 3: Multimodal, multilingual, long context open models.
\newblock Technical report, Google, 2025.

\bibitem[Dubey et~al.(2024)Dubey, Jauhri, Pandey, Kadian, Al-Dahle, Letak, Mathur, Schelten, Yang, Fan, et~al.]{dubey2024llama}
Abhimanyu Dubey, Abhinav Jauhri, Abhinav Pandey, Abhishek Kadian, Ahmad Al-Dahle, Aiesha Letak, Akhil Mathur, Alan Schelten, Amy Yang, Angela Fan, et~al.
\newblock The llama 3 herd of models.
\newblock \emph{arXiv preprint arXiv:2407.21783}, 2024.

\bibitem[Fernandez et~al.(2023)Fernandez, Chaffin, Tit, Chappelier, and Furon]{fernandez2023three}
Pierre Fernandez, Antoine Chaffin, Karim Tit, Vivien Chappelier, and Teddy Furon.
\newblock Three bricks to consolidate watermarks for large language models.
\newblock \emph{2023 IEEE International Workshop on Information Forensics and Security (WIFS)}, 2023.

\bibitem[Fu et~al.(2024)Fu, Xiong, and Dong]{fu2024watermarking}
Yu Fu, Deyi Xiong, and Yue Dong.
\newblock Watermarking conditional text generation for ai detection: Unveiling challenges and a semantic-aware watermark remedy.
\newblock In \emph{Proceedings of the AAAI Conference on Artificial Intelligence}, pages 18003--18011, 2024.

\bibitem[Giboulot and Furon(2024)]{giboulot2024watermax}
Eva Giboulot and Teddy Furon.
\newblock Watermax: breaking the llm watermark detectability-robustness-quality trade-off.
\newblock \emph{arXiv preprint arXiv:2403.04808}, 2024.

\bibitem[Holtzman et~al.(2019)Holtzman, Buys, Du, Forbes, and Choi]{holtzman2019curious}
Ari Holtzman, Jan Buys, Li Du, Maxwell Forbes, and Yejin Choi.
\newblock The curious case of neural text degeneration.
\newblock \emph{arXiv preprint arXiv:1904.09751}, 2019.

\bibitem[Hou et~al.(2023)Hou, Zhang, He, Wang, Chuang, Wang, Shen, Van~Durme, Khashabi, and Tsvetkov]{hou2023semstamp}
Abe~Bohan Hou, Jingyu Zhang, Tianxing He, Yichen Wang, Yung-Sung Chuang, Hongwei Wang, Lingfeng Shen, Benjamin Van~Durme, Daniel Khashabi, and Yulia Tsvetkov.
\newblock Semstamp: A semantic watermark with paraphrastic robustness for text generation.
\newblock \emph{arXiv preprint arXiv:2310.03991}, 2023.

\bibitem[Hou et~al.(2024)Hou, Zhang, Wang, Khashabi, and He]{hou2024k}
Abe~Bohan Hou, Jingyu Zhang, Yichen Wang, Daniel Khashabi, and Tianxing He.
\newblock k-semstamp: A clustering-based semantic watermark for detection of machine-generated text.
\newblock \emph{arXiv preprint arXiv:2402.11399}, 2024.

\bibitem[Huang et~al.(2023)Huang, Zhu, Zhu, Lee, Jiao, and Jordan]{huang2023optimal}
Baihe Huang, Banghua Zhu, Hanlin Zhu, Jason~D. Lee, Jiantao Jiao, and Michael~I. Jordan.
\newblock Towards optimal statistical watermarking, 2023.

\bibitem[Jiang et~al.(2023)Jiang, Sablayrolles, Mensch, Bamford, Devendra, de~las Casas, Bressand, Lengyel, Lample, Saulnier, Lavaud, Lachaux, Stock, Scao, Lavril, Wang, Lacroix, and Sayed]{jiang2023mistral}
Albert~Q. Jiang, Alexandre Sablayrolles, Arthur Mensch, Chris Bamford, Singh~Chaplot Devendra, Diego de~las Casas, Florian Bressand, Gianna Lengyel, Guillaume Lample, Lucile Saulnier, Lélio Lavaud, Marie-Anne Lachaux, Pierre Stock, Teven~Le Scao, Thibaut Lavril, Thomas Wang, Timothée Lacroix, and William~El Sayed.
\newblock Mistral 7b.
\newblock \emph{arXiv preprint arXiv:2310.06825}, 2023.

\bibitem[Jovanovi{\'c} et~al.(2025)Jovanovi{\'c}, Staab, Baader, and Vechev]{jovanovic2025ward}
Nikola Jovanovi{\'c}, Robin Staab, Maximilian Baader, and Martin Vechev.
\newblock Ward: Provable rag dataset inference via llm watermarks.
\newblock \emph{ICLR}, 2025.

\bibitem[Kirchenbauer et~al.(2023{\natexlab{a}})Kirchenbauer, Geiping, Wen, Katz, Miers, and Goldstein]{kirchenbauer2023watermark}
John Kirchenbauer, Jonas Geiping, Yuxin Wen, Jonathan Katz, Ian Miers, and Tom Goldstein.
\newblock A watermark for large language models.
\newblock \emph{arXiv preprint arXiv:2301.10226}, 2023{\natexlab{a}}.

\bibitem[Kirchenbauer et~al.(2023{\natexlab{b}})Kirchenbauer, Geiping, Wen, Shu, Saifullah, Kong, Fernando, Saha, Goldblum, and Goldstein]{kirchenbauer2023reliability}
John Kirchenbauer, Jonas Geiping, Yuxin Wen, Manli Shu, Khalid Saifullah, Kezhi Kong, Kasun Fernando, Aniruddha Saha, Micah Goldblum, and Tom Goldstein.
\newblock On the reliability of watermarks for large language models, 2023{\natexlab{b}}.

\bibitem[Kuditipudi et~al.(2023)Kuditipudi, Thickstun, Hashimoto, and Liang]{kuditipudi2023robust}
Rohith Kuditipudi, John Thickstun, Tatsunori Hashimoto, and Percy Liang.
\newblock Robust distortion-free watermarks for language models.
\newblock \emph{arXiv preprint arXiv:2307.15593}, 2023.

\bibitem[Lau et~al.(2024)Lau, Niu, Dao, Chen, Foo, and Low]{lau2024waterfall}
Gregory Kang~Ruey Lau, Xinyuan Niu, Hieu Dao, Jiangwei Chen, Chuan-Sheng Foo, and Bryan Kian~Hsiang Low.
\newblock Waterfall: Framework for robust and scalable text watermarking.
\newblock In \emph{ICML 2024 Workshop on Foundation Models in the Wild}, 2024.

\bibitem[Lee et~al.(2023)Lee, Hong, Ahn, Hong, Lee, Yun, Shin, and Kim]{lee2023wrote}
Taehyun Lee, Seokhee Hong, Jaewoo Ahn, Ilgee Hong, Hwaran Lee, Sangdoo Yun, Jamin Shin, and Gunhee Kim.
\newblock Who wrote this code? watermarking for code generation.
\newblock \emph{arXiv preprint arXiv:2305.15060}, 2023.

\bibitem[Liu et~al.(2023)Liu, Pan, Hu, Meng, and Wen]{liu2023semantic}
Aiwei Liu, Leyi Pan, Xuming Hu, Shiao Meng, and Lijie Wen.
\newblock A semantic invariant robust watermark for large language models.
\newblock \emph{arXiv preprint arXiv:2310.06356}, 2023.

\bibitem[Liu and Bu(2024)]{liu2024adaptive}
Yepeng Liu and Yuheng Bu.
\newblock Adaptive text watermark for large language models.
\newblock \emph{arXiv preprint arXiv:2401.13927}, 2024.

\bibitem[Meral et~al.(2009)Meral, Sankur, {\"O}zsoy, G{\"u}ng{\"o}r, and Sevin{\c{c}}]{meral2009natural}
Hasan~Mesut Meral, B{\"u}lent Sankur, A~Sumru {\"O}zsoy, Tunga G{\"u}ng{\"o}r, and Emre Sevin{\c{c}}.
\newblock Natural language watermarking via morphosyntactic alterations.
\newblock \emph{Computer Speech \& Language}, 23\penalty0 (1):\penalty0 107--125, 2009.

\bibitem[OpenAI(2022)]{chatgpt2022}
OpenAI.
\newblock {ChatGPT}: Optimizing language models for dialogue., 2022.

\bibitem[Pan et~al.(2024)Pan, Liu, He, Gao, Zhao, Lu, Zhou, Liu, Hu, Wen, et~al.]{pan2024markllm}
Leyi Pan, Aiwei Liu, Zhiwei He, Zitian Gao, Xuandong Zhao, Yijian Lu, Binglin Zhou, Shuliang Liu, Xuming Hu, Lijie Wen, et~al.
\newblock Markllm: An open-source toolkit for llm watermarking.
\newblock \emph{arXiv preprint arXiv:2405.10051}, 2024.

\bibitem[Piet et~al.(2023)Piet, Sitawarin, Fang, Mu, and Wagner]{piet2023mark}
Julien Piet, Chawin Sitawarin, Vivian Fang, Norman Mu, and David Wagner.
\newblock Mark my words: Analyzing and evaluating language model watermarks.
\newblock \emph{arXiv preprint arXiv:2312.00273}, 2023.

\bibitem[{Project Gutenberg}(2025)]{projectgutenberg}
{Project Gutenberg}.
\newblock Project gutenberg, 2025.
\newblock Accessed: 2025-11-15.

\bibitem[Qiang et~al.(2023)Qiang, Zhu, Li, Zhu, Yuan, and Wu]{qiang2023natural}
Jipeng Qiang, Shiyu Zhu, Yun Li, Yi Zhu, Yunhao Yuan, and Xindong Wu.
\newblock Natural language watermarking via paraphraser-based lexical substitution.
\newblock \emph{Artificial Intelligence}, 317:\penalty0 103859, 2023.

\bibitem[Qu et~al.(2024)Qu, Yin, He, Zou, Tao, Jia, and Zhang]{qu2024provably}
Wenjie Qu, Dong Yin, Zixin He, Wei Zou, Tianyang Tao, Jinyuan Jia, and Jiaheng Zhang.
\newblock Provably robust multi-bit watermarking for ai-generated text via error correction code.
\newblock \emph{arXiv preprint arXiv:2401.16820}, 2024.

\bibitem[Rastogi et~al.(2025)Rastogi, Maini, and Pruthi]{rastogi2025stamp}
Saksham Rastogi, Pratyush Maini, and Danish Pruthi.
\newblock Stamp your content: Proving dataset membership via watermarked rephrasings.
\newblock \emph{arXiv preprint arXiv:2504.13416}, 2025.

\bibitem[Sander et~al.(2024)Sander, Fernandez, Durmus, Douze, and Furon]{sander2024watermarking}
Tom Sander, Pierre Fernandez, Alain Durmus, Matthijs Douze, and Teddy Furon.
\newblock Watermarking makes language models radioactive.
\newblock \emph{NeurIPS}, 2024.

\bibitem[Sander et~al.(2025)Sander, Fernandez, Mahloujifar, Durmus, and Guo]{sander2025detecting}
Tom Sander, Pierre Fernandez, Saeed Mahloujifar, Alain Durmus, and Chuan Guo.
\newblock Detecting benchmark contamination through watermarking.
\newblock \emph{arXiv preprint arXiv:2502.17259}, 2025.

\bibitem[Shirali-Shahreza and Shirali-Shahreza(2008)]{shirali2008new}
M~Hassan Shirali-Shahreza and Mohammad Shirali-Shahreza.
\newblock A new synonym text steganography.
\newblock In \emph{2008 international conference on intelligent information hiding and multimedia signal processing}, pages 1524--1526. IEEE, 2008.

\bibitem[Sutskever et~al.(2014)Sutskever, Vinyals, and Le]{sutskever2014sequence}
Ilya Sutskever, Oriol Vinyals, and Quoc~V Le.
\newblock Sequence to sequence learning with neural networks.
\newblock In \emph{NeurIPS}, 2014.

\bibitem[Team(2024)]{qwen25}
Qwen Team.
\newblock Qwen2.5 technical report.
\newblock \emph{arXiv preprint arXiv:2409.12117}, 2024.

\bibitem[Topkara et~al.(2005)Topkara, Taskiran, and Delp~III]{topkara2005natural}
Mercan Topkara, Cuneyt~M Taskiran, and Edward~J Delp~III.
\newblock Natural language watermarking.
\newblock In \emph{Security, Steganography, and Watermarking of Multimedia Contents VII}, pages 441--452. SPIE, 2005.

\bibitem[Topkara et~al.(2006{\natexlab{a}})Topkara, Riccardi, Hakkani-T{\"u}r, and Atallah]{topkara2006natural}
Mercan Topkara, Giuseppe Riccardi, Dilek Hakkani-T{\"u}r, and Mikhail~J Atallah.
\newblock Natural language watermarking: Challenges in building a practical system.
\newblock In \emph{Security, Steganography, and Watermarking of Multimedia Contents VIII}, pages 106--117. SPIE, 2006{\natexlab{a}}.

\bibitem[Topkara et~al.(2006{\natexlab{b}})Topkara, Topkara, and Atallah]{topkara2006words}
Mercan Topkara, Umut Topkara, and Mikhail~J Atallah.
\newblock Words are not enough: sentence level natural language watermarking.
\newblock In \emph{Proceedings of the 4th ACM international workshop on Contents protection and security}, pages 37--46, 2006{\natexlab{b}}.

\bibitem[Topkara et~al.(2006{\natexlab{c}})Topkara, Topkara, and Atallah]{topkara2006hiding}
Umut Topkara, Mercan Topkara, and Mikhail~J Atallah.
\newblock The hiding virtues of ambiguity: quantifiably resilient watermarking of natural language text through synonym substitutions.
\newblock In \emph{Proceedings of the 8th workshop on Multimedia and security}, pages 164--174, 2006{\natexlab{c}}.

\bibitem[Ueoka et~al.(2021)Ueoka, Murawaki, and Kurohashi]{ueoka2021frustratingly}
Honai Ueoka, Yugo Murawaki, and Sadao Kurohashi.
\newblock Frustratingly easy edit-based linguistic steganography with a masked language model.
\newblock \emph{arXiv preprint arXiv:2104.09833}, 2021.

\bibitem[Venugopal et~al.(2011)Venugopal, Uszkoreit, Talbot, Och, and Ganitkevitch]{venugopal2011watermarking}
Ashish Venugopal, Jakob Uszkoreit, David Talbot, Franz~Josef Och, and Juri Ganitkevitch.
\newblock Watermarking the outputs of structured prediction with an application in statistical machine translation.
\newblock In \emph{Proceedings of the 2011 Conference on Empirical Methods in Natural Language Processing}, pages 1363--1372, 2011.

\bibitem[Wang et~al.(2025)Wang, Gu, Wu, and Yang]{wang2025morphmark}
Zongqi Wang, Tianle Gu, Baoyuan Wu, and Yujiu Yang.
\newblock Morphmark: Flexible adaptive watermarking for large language models.
\newblock \emph{arXiv preprint arXiv:2505.11541}, 2025.

\bibitem[Weidinger et~al.(2022)Weidinger, Uesato, Rauh, Griffin, Huang, Mellor, Glaese, Cheng, Balle, Kasirzadeh, et~al.]{weidinger2022taxonomy}
Laura Weidinger, Jonathan Uesato, Maribeth Rauh, Conor Griffin, Po-Sen Huang, John Mellor, Amelia Glaese, Myra Cheng, Borja Balle, Atoosa Kasirzadeh, et~al.
\newblock Taxonomy of risks posed by language models.
\newblock In \emph{2022 ACM Conference on Fairness, Accountability, and Transparency}, pages 214--229, 2022.

\bibitem[Wilson and Ker(2016)]{wilson2016avoiding}
Alex Wilson and Andrew~D Ker.
\newblock Avoiding detection on twitter: embedding strategies for linguistic steganography.
\newblock \emph{Electronic Imaging}, 28:\penalty0 1--9, 2016.

\bibitem[Winstein(1998)]{winstein1998lexical}
Keith Winstein.
\newblock Lexical steganography through adaptive modulation of the word choice hash.
\newblock \emph{Unpublished. http://www. imsa. edu/\~{} keithw/tlex}, 1998.

\bibitem[Wu et~al.(2023)Wu, Hu, Zhang, and Huang]{wu2023dipmark}
Yihan Wu, Zhengmian Hu, Hongyang Zhang, and Heng Huang.
\newblock Dipmark: A stealthy, efficient and resilient watermark for large language models.
\newblock \emph{arXiv preprint arXiv:2310.07710}, 2023.

\bibitem[Xiang et~al.(2017)Xiang, Wang, Yang, and Liu]{xiang2017novel}
Lingyun Xiang, Xinhui Wang, Chunfang Yang, and Peng Liu.
\newblock A novel linguistic steganography based on synonym run-length encoding.
\newblock \emph{IEICE transactions on Information and Systems}, 100\penalty0 (2):\penalty0 313--322, 2017.

\bibitem[Xu et~al.(2024)Xu, Jia, Yao, Liu, and Li]{xu2024robust}
Xiaojun Xu, Jinghan Jia, Yuanshun Yao, Yang Liu, and Hang Li.
\newblock Robust multi-bit text watermark with llm-based paraphrasers.
\newblock \emph{arXiv preprint arXiv:2412.03123}, 2024.

\bibitem[Yoo et~al.(2023{\natexlab{a}})Yoo, Ahn, Jang, and Kwak]{yoo2023robust}
KiYoon Yoo, Wonhyuk Ahn, Jiho Jang, and Nojun Kwak.
\newblock Robust multi-bit natural language watermarking through invariant features.
\newblock \emph{arXiv preprint arXiv:2305.01904}, 2023{\natexlab{a}}.

\bibitem[Yoo et~al.(2023{\natexlab{b}})Yoo, Ahn, and Kwak]{yoo2023advancing}
KiYoon Yoo, Wonhyuk Ahn, and Nojun Kwak.
\newblock Advancing beyond identification: Multi-bit watermark for language models.
\newblock \emph{arXiv preprint arXiv:2308.00221}, 2023{\natexlab{b}}.

\bibitem[Yoo et~al.(2024)Yoo, Ahn, and Kwak]{yoo2024advancing}
KiYoon Yoo, Wonhyuk Ahn, and Nojun Kwak.
\newblock Advancing beyond identification: Multi-bit watermark for large language models.
\newblock In \emph{Proceedings of the 2024 Conference of the North American Chapter of the Association for Computational Linguistics: Human Language Technologies (Volume 1: Long Papers)}, pages 4031--4055, 2024.

\bibitem[Zhang et~al.(2025)Zhang, Chen, Yang, Mai, Huang, and Pang]{zhang2025leave}
Jingqi Zhang, Ruibo Chen, Yingqing Yang, Peihua Mai, Heng Huang, and Yan Pang.
\newblock Leave no trace: Black-box detection of copyrighted dataset usage in large language models via watermarking.
\newblock \emph{arXiv preprint arXiv:2510.02962}, 2025.

\bibitem[Zhang et~al.(2024)Zhang, Hussain, Neekhara, and Koushanfar]{zhang2024remark}
Ruisi Zhang, Shehzeen~Samarah Hussain, Paarth Neekhara, and Farinaz Koushanfar.
\newblock $\{$REMARK-LLM$\}$: A robust and efficient watermarking framework for generative large language models.
\newblock In \emph{33rd USENIX Security Symposium (USENIX Security 24)}, pages 1813--1830, 2024.

\bibitem[Zhang et~al.(2020)Zhang, Kishore, Wu, Weinberger, and Artzi]{zhang2020bertscore}
Tianyi Zhang, Varsha Kishore, Felix Wu, Kilian~Q Weinberger, and Yoav Artzi.
\newblock Bertscore: Evaluating text generation with bert.
\newblock In \emph{International Conference on Learning Representations}, 2020.

\bibitem[Zhao et~al.(2023)Zhao, Ananth, Li, and Wang]{zhao2023provable}
Xuandong Zhao, Prabhanjan Ananth, Lei Li, and Yu-Xiang Wang.
\newblock Provable robust watermarking for ai-generated text.
\newblock \emph{arXiv preprint arXiv:2306.17439}, 2023.

\end{thebibliography}

\clearpage

\beginappendix

\section{Additional Details on LLM Watermarking}

\subsection{Hash Function}\label{app:hash}

All watermarking schemes in this work rely on a pseudorandom function (PRF) to deterministically map how watermark windows should influence the next-token selection (e.g., green/red classification or Gumbel noise values).
The PRF takes as input the candidate token $x$, a context window $\mathbf{w} = (w_1, \ldots, w_k)$ of $k$ token IDs, and the secret key $\sk$ (all of them are integers), and outputs a random integer in $[0, M)$.
We compute:
\begin{align}
h'(x, \mathbf{w}, \sk) & = \left( p_2 \cdot x + \sum_{i=1}^{k} w_i \cdot q_i + p_3 \cdot \sk \right) \cdot p_4, \\
h (x, \mathbf{w}, \sk) & = \text{XORShift}(h'(x, \mathbf{w}, \sk)) \mod M,
\end{align}
where $q_1, \ldots, q_k$ are distinct large primes (to ensure that different orderings of the same tokens produce different tokens), and $p_2, p_3, p_4$ are additional primes. 
The first result $h'$ undergoes XOR-shift for better bit dispersion: $h = (h' \cdot p_{\text{mix}}) \oplus ( (h' \cdot p_{\text{mix}}) \gg s)$, where $p_{\text{mix}}$ is a mixing prime and $s$ is a shift constant. 
Finally, $h \mod M$ yields an integer in $[0, M)$.

For uniform PRF output in $[0,1)$, we divide by $M$.
For binary (green/red) classification, we threshold at $\gamma$: a token is ``green'' if $h/M < \gamma$.
This construction ensures that small changes in any input (window, token, or key) produce uncorrelated outputs, satisfying the pseudo-randomness requirements for watermark security.
In practice, the implementation allows for selectively scoring some tokens without needing to generate the full green/red list, which can speed up computations for large vocabularies, especially for SynthID-text, which uses multiple lists for different $g$-values.

\subsection{Details on Watermark Schemes}\label{app:method_overview}\label{app:related_work}

We describe below the main watermarking schemes we evaluate.

\paragraph{\textbf{Green-list/red-list.}}~\citet{kirchenbauer2023watermark,kirchenbauer2023reliability} modify the logit vector during next-token generation based on the watermark window of $k$ previous tokens and the private key $\sk$.
Note that the case where $k=0$ corresponds to the work of \citet{zhao2023provable}, but we found that it empirically breaks theoretical assumptions and often leads to degenerate text; therefore, we use $k \geq 1$ in our experiments.
For each token $v \in \V$, the pseudorandom function $\mathrm{PRF}(\cdot)$ outputs a value in $[0,1)$ from the token ID $v$, the context window, and $\sk$.
A token is classified as ``green'' if $\mathrm{PRF}(v, \mathbf{w}, \sk) < \gamma$, where $\gamma \in [0,1]$ is the expected proportion of green tokens under the null hypothesis (typically $\gamma=0.5$).
Logits of green tokens are incremented by $\delta$ to increase their sampling probability:
\begin{equation}
\tilde{\ell}_v = 
\begin{cases}
\ell_v + \delta & \text{if } v \in \text{GreenList} \\
\ell_v & \text{otherwise}
\end{cases}
\end{equation}
Detection involves repeating the greenlist computation for each token of a text, incrementing a score by 1 if the token is in the greenlist, and performing a statistical test on the cumulative score.
Under the null hypothesis $\H_0$ ``the text is not watermarked with that scheme and private key $\sk$'', this score follows a binomial distribution.
A simple binomial test thus provides a $p$-value: the probability of observing at least as many green tokens as observed, if the text was not watermarked.

\paragraph{\textbf{SWEET.}}
\citet{lee2023wrote} apply the Green-red list watermark only to high-entropy tokens. 
The intuition is that low-entropy tokens (where the model is confident) contribute little signal but can degrade quality if biased. 
At generation time, tokens with entropy below a threshold are sampled without watermark bias. 
Specifically, a token is considered high-entropy if $H(\mathbf{p}) = -\sum_{v \in \V} p_v \log p_v > \tau$, where $\tau$ is a predefined threshold.
At detection time, these tokens are similarly excluded from scoring, improving signal-to-noise ratio. 
Note that this filtering can also be applied only at detection time, and on top of other schemes than green-red (as we do in our experiments).

\paragraph{\textbf{MorphMark.}}
\citet{wang2025morphmark} adaptively adjust watermark strength based on context. 
Let $P_G = \sum_{v \in \text{GreenList}} p_v$ be the total probability mass on green tokens before watermarking. 
If $P_G \leq p_0$ (a threshold, e.g., $p_0=0.15$), no watermark is applied to preserve quality. 
Otherwise, an adaptive boost factor $r = \min (\kappa P_G, 1)$ is computed, where $\kappa$ controls the watermark strength. 
Probabilities are then adjusted:
\begin{equation}
\hat{p}_v = 
\begin{cases}
p_v \cdot \left(1 + \frac{r(1-P_G)}{P_G}\right) & \text{if } v \in \text{GreenList} \\
p_v \cdot (1 - r) & \text{otherwise}
\end{cases}
\end{equation}
This ensures stronger watermarking when the green list is already favorable, minimizing quality degradation.

\paragraph{\textbf{DiPMark.}}
\citet{wu2023dipmark} introduce a variant of Green-red watermarks that is distortion-free. 
The method uses a pseudorandom permutation $\pi$ (seeded by the context window and $\sk$) to reorder tokens. 
After permutation, the cumulative probability distribution is modified as follows: 
tokens in the interval $[0, 1-\alpha]$ have their probability set to zero, tokens in $[1-\alpha, \alpha]$ remain unchanged, and tokens in $[\alpha, 1]$ have their probability doubled (then renormalized). 
This creates a detectable bias while preserving the original distribution's average over the randomness of $\pi$.

\paragraph{\textbf{Gumbel-max.}}
\citet{aaronson2023watermarking} alternatively leverage the ``Gumbel-max trick''.
After applying temperature scaling and optional top-$k$ or top-$p$ filtering to obtain a probability vector $\mathbf{p}$, the next token is selected as:
\begin{equation}
x_{t} = \arg\max_{v \in \V} r_v^{1/p_v}
\end{equation}
where $r_v \sim \text{Uniform}(0,1)$ is i.i.d. noise (which is equivalent to sampling from $\mathbf{p}$).
The watermark intervenes by replacing the purely random $r_v$ with pseudorandom values $r_v = \mathrm{PRF}(v, \mathbf{w}, \sk) \in [0,1)$ generated by the PRF from the token ID $v$, the context window, and $\sk$, for each $v \in \V$.
Consequently, for a fixed context, the noise vector is deterministic.
Detection is performed by recomputing the PRF output $r_{x_t} = \mathrm{PRF}(x_t, \mathbf{w}_t, \sk)$ for each observed token $x_t$ and computing the cumulative score:
\begin{equation}
S = \sum_{t=1}^{N} -\log(1 - r_{x_t})
\end{equation}
where $N$ is the number of tokens in the text.
Under $\H_0$ ``the text is not watermarked with that scheme and private key $\sk$'', $S$ follows a $\Gamma(N,1)$ distribution.
The $p$-value of a test associated with score $S$ reads: $\text{$p$-value}(S) = \Gamma(N,S)/\Gamma(N)$, where $\Gamma$ is the upper incomplete gamma function.

\paragraph{\textbf{SynthID-text.}}
\citet{dathathri2024scalable} propose SynthID Text, which employs a tournament-based sampling strategy.
The method uses $m$ scoring functions $g_1, \ldots, g_m$, where each $g_i$ maps a token to a pseudorandom value in $\{0,1\}$ based on the context window and secret key\footnote{In our implementation, each $g$-function corresponds to a distinct green/red list partition: we compute them by incrementing the token ID by $i$ for each depth $i$ of the tournament.
Since the hash function exhibits no correlation for incremented token IDs, this effectively simulates independent green/red lists per tournament layer.}.
Unlike the additive bias of the ``Green-red'' list method, this approach samples $2^m$ candidate tokens from the model's distribution and randomly organizes them into a tournament where winners are determined by the $g$-values.
Detection relies on a statistical hypothesis test using the mean $g$-value of the observed tokens as the test statistic.
Under the null hypothesis $\H_0$ (unwatermarked text), the $g$-values are independent and identically distributed (e.g., uniform or Bernoulli).
This scheme is shown to outperform Green-red approaches because the tournament depth $m$ allows for a more favorable trade-off between detectability and text quality (perplexity), particularly in low-entropy settings where standard additive biases often degrade coherence.

Scoring Functions: Mean vs. Weighted Mean.
While the standard detection method computes a simple unweighted mean of the $g$-values across all tokens and tournament layers (effectively a normalized sum analogous to a binomial test), \citet{dathathri2024scalable} demonstrate that this is theoretically suboptimal for multi-layer tournaments.
The authors observe that the amount of ``watermarking evidence'' embedded by the tournament is not uniform across layers; specifically, the strength of the watermark diminishes as the tournament depth increases because each successive layer operates on a subset of candidates with reduced entropy.
Consequently, an unweighted mean dilutes the strong signal from early layers with the weaker signal from deeper layers.
To address this, SynthID-text utilizes a \textit{weighted mean score}, which assigns decreasing weights $w_l$ to the $g$-values of the $l$-th tournament layer.
By emphasizing the earlier layers where the statistical signature is most robust, the weighted scoring function improves the signal-to-noise ratio of the test statistic, yielding higher detection accuracy (true positive rate) for a fixed false positive rate compared to the classical unweighted approach.

\subsection{Choice of the Secret Key and Statistical Correctness}

\paragraph{How to choose the secret key?}
As explained in~\autoref{sec:method}, valid detection requires $p$-values to be uniformly distributed under $\H_0$.
While this holds in expectation over all possible keys, fixing a specific key $\sk$ creates preferences for certain $(k+1)$-grams.
If these patterns align with natural language statistics, the expected green fraction under $\H_0$ shifts slightly (e.g., from $0.5$ to $0.505$), causing $p$-values to collapse toward zero on long texts and inflating false positive rates.

We address this by testing 50 candidate keys for each combination of tokenizer, watermarking method, and relevant hyperparameters.
Each key is evaluated on 100 non-watermarked texts of 800 characters, and we select the key that minimizes deviation from $U(0,1)$ as measured by the Kolmogorov-Smirnov (KS) statistic.
This ensures our results reflect intrinsic scheme performance rather than key-specific artifacts.
Detailed statistics are provided in~\autoref{tab:h0_key_sensitivity}.

\begin{table*}[t!]
    \centering
    \scriptsize
    \caption{Key sensitivity analysis across models and watermarking schemes. We evaluate multiple random keys per configuration on 100 non-watermarked texts. ``SynthID (W)'' denotes the Weighted variant which utilizes a different scoring statistic. The \textbf{Best Key} is selected based on the highest $p$-value from the Kolmogorov-Smirnov (KS) test for uniformity under $\H_0$.}
    \label{tab:h0_key_sensitivity}
    \begin{tabular}{ll c c | c c c}
        \toprule
        & & \multicolumn{2}{c}{\textbf{Aggregated Statistics (All Keys)}} & \multicolumn{3}{c}{\textbf{Selected ``Best'' Key}} \\
        \cmidrule(lr){3-4} \cmidrule(lr){5-7}
        \textbf{Model} & \textbf{Method} & \textbf{Avg. $p$-val} & \textbf{$\sigma_{\text{keys}}$} & \textbf{Key ID} & \textbf{Avg. $p$-val} & \textbf{KS $p$} \\
        \midrule

        \multirow{10}{*}{\textbf{Qwen 2.5}} 
        & DipMark & 0.501 & 0.033 & \#596061 & 0.493 & 0.99 \\
        & Green-red & 0.501 & 0.033 & \#596061 & 0.493 & 0.99 \\
        & MorphMark & 0.501 & 0.033 & \#596061 & 0.493 & 0.99 \\
        & Gumbel-max & 0.497 & 0.031 & \#2345 & 0.499 & 0.99 \\
        & SynthID ($d=10$) & 0.498 & 0.035 & \#606 & 0.494 & 0.95 \\
        & SynthID ($d=20$) & 0.488 & 0.033 & \#753 & 0.506 & 0.99 \\
        & SynthID ($d=30$) & 0.488 & 0.031 & \#1357 & 0.493 & 0.97 \\
        & SynthID (W) ($d=10$) & 0.498 & 0.030 & \#323334 & 0.506 & 0.98 \\
        & SynthID (W) ($d=20$) & 0.493 & 0.029 & \#3452 & 0.498 & 0.99 \\
        & SynthID (W) ($d=30$) & 0.490 & 0.032 & \#505152 & 0.509 & 1.00 \\
        \midrule

        \multirow{10}{*}{\textbf{Llama 3}} 
        & DipMark & 0.508 & 0.037 & \#323334 & 0.505 & 0.95 \\
        & Green-red & 0.508 & 0.037 & \#323334 & 0.505 & 0.95 \\
        & MorphMark & 0.508 & 0.037 & \#323334 & 0.505 & 0.95 \\
        & Gumbel-max & 0.498 & 0.030 & \#6780 & 0.506 & 1.00 \\
        & SynthID ($d=10$) & 0.507 & 0.035 & \#656667 & 0.504 & 0.99 \\
        & SynthID ($d=20$) & 0.497 & 0.035 & \#626364 & 0.493 & 0.99 \\
        & SynthID ($d=30$) & 0.503 & 0.037 & \#686970 & 0.499 & 0.99 \\
        & SynthID (W) ($d=10$) & 0.504 & 0.036 & \#888 & 0.507 & 0.96 \\
        & SynthID (W) ($d=20$) & 0.500 & 0.034 & \#42 & 0.504 & 1.00 \\
        & SynthID (W) ($d=30$) & 0.502 & 0.036 & \#333 & 0.490 & 0.98 \\
        \midrule

        \multirow{10}{*}{\textbf{Gemma 3}} 
        & DipMark & 0.501 & 0.035 & \#8907 & 0.488 & 0.96 \\
        & Green-red & 0.501 & 0.035 & \#8907 & 0.488 & 0.96 \\
        & MorphMark & 0.501 & 0.035 & \#8907 & 0.488 & 0.96 \\
        & Gumbel-max & 0.498 & 0.034 & \#6789 & 0.506 & 0.97 \\
        & SynthID ($d=10$) & 0.503 & 0.041 & \#656667 & 0.504 & 0.98 \\
        & SynthID ($d=20$) & 0.496 & 0.036 & \#323334 & 0.488 & 0.99 \\
        & SynthID ($d=30$) & 0.502 & 0.036 & \#131415 & 0.495 & 0.93 \\
        & SynthID (W) ($d=10$) & 0.501 & 0.042 & \#1234 & 0.519 & 0.97 \\
        & SynthID (W) ($d=20$) & 0.492 & 0.036 & \#258 & 0.501 & 0.99 \\
        & SynthID (W) ($d=30$) & 0.503 & 0.039 & \#222324 & 0.498 & 0.98 \\
        \midrule

        \multirow{10}{*}{\textbf{SmolLM2}} 
        & DipMark & 0.496 & 0.034 & \#789 & 0.506 & 1.00 \\
        & Green-red & 0.496 & 0.034 & \#789 & 0.506 & 1.00 \\
        & MorphMark & 0.496 & 0.034 & \#789 & 0.506 & 1.00 \\
        & Gumbel-max & 0.507 & 0.038 & \#252627 & 0.498 & 1.00 \\
        & SynthID ($d=10$) & 0.506 & 0.044 & \#535455 & 0.501 & 0.98 \\
        & SynthID ($d=20$) & 0.495 & 0.041 & \#369 & 0.499 & 1.00 \\
        & SynthID ($d=30$) & 0.497 & 0.043 & \#444 & 0.500 & 0.99 \\
        & SynthID (W) ($d=10$) & 0.502 & 0.043 & \#707 & 0.509 & 0.98 \\
        & SynthID (W) ($d=20$) & 0.499 & 0.041 & \#369 & 0.499 & 1.00 \\
        & SynthID (W) ($d=30$) & 0.501 & 0.044 & \#252627 & 0.507 & 0.98 \\

        \bottomrule
    \end{tabular}
\end{table*}

\subsection{Details on Radioactivity}\label{app:radioactivity_details}

\paragraph{\textbf{The Radioactivity Test Protocol.}}

To formally test for watermarked dataset radioactivity, we detail the methodology from~\citet{sander2024watermarking,sander2025detecting} for the Green-list/Red-list scheme.
The core idea is to repurpose the standard watermark detection test (normally applied to observed text tokens) and instead apply it to the \textit{predicted} tokens of the suspect model.
This allows us to determine whether the model is familiar with the watermark, thereby providing evidence that the model was exposed to the watermark during training.

\paragraph{\textbf{Teacher Forcing Setup.}}

We feed the \textit{watermarked} text $\mathbf{y}$ into the suspect model $f_\theta$ using a teacher-forcing setup. Let $\hat{y}_t$ denote the top-1 prediction of the suspect model at step $t$, given the watermarked prefix $y_{<t}$:

\begin{equation}
    \hat{y}_t = \underset{v \in \mathcal{V}}{\mathrm{argmax}} \ P_\theta(v \mid y_{<t}).
\end{equation}

\paragraph{\textbf{Test Statistic.}}

We define the radioactivity score $S$ as the empirical proportion of the suspect model's predictions that fall into the Green-list:

\begin{equation}
    S = \frac{1}{|\mathcal{U}|} \sum_{t \in \mathcal{U}} \mathbb{I}\bigl( h(\hat{y}_t, y_{t-k:t-1}, \sk) < \gamma \bigr),
\end{equation}

where $\sk$ is the secret key and $\gamma$ is the expected green ratio (typically $0.5$).

\paragraph{\textbf{Deduplication.}}
To ensure statistical independence, we must account for repeated n-grams in the context that could influence the model to reproduce them.
In particular, if an n-gram in context is green, the suspect model might repeat it due to context copying rather than watermark radioactivity.
We filter the indices $t$ to form a set $\mathcal{U}$ such that each watermark window $y_{t-k:t-1}$ is only scored once, which fully removes this issue~\citep{sander2024watermarking}.
Under the null hypothesis $\mathcal{H}_0$ (i.e., the suspect model is unaware of $\sk$), the count of green predictions should now follow a binomial distribution, as the suspect model should exhibit no preference toward green or red tokens:

\begin{equation}
    K \sim \text{Binomial}(|\mathcal{U}|, \gamma).
\end{equation}

This formulation allows for the computation of an exact $p$-value.

\paragraph{\textbf{WaterMax Is Not Radioactive.}}

For the score to truly follow a known binomial distribution under $\mathcal{H}_0$ (``the suspect model has never been in contact with the watermark''), the watermark bias must be applied at the \textit{token level}.
However, WaterMax selects the final sequence $x^*$ from a set of candidates $\mathcal{C}$ by maximizing the number of green tokens globally \emph{by chance}.
Consequently, an innocent model $f_\theta$ that shares a similar language distribution with the generator $P_{\text{gen}}$ will find that its optimal next token $\hat{y}_t$ is green significantly more often than $\gamma$, purely due to this selection bias.

We verified this experimentally by running the radioactivity detection test on approximately 100k watermarked tokens generated via WaterMax, Green-list/Red-list random sampling, and Green-list/Red-list beam search with biased scoring (after rephrasing 300 excerpts from Dickens as detailed in~\autoref{subsec:exp_set_up}).
As expected, the radioactivity test yields a $p$-value of 0.93 for Green-list/Red-list random sampling and 0.76 for Green-list/Red-list beam search with biased scoring, \emph{but $1.0 \times 10^{-6}$ for WaterMax}.

\section{Additional Text Samples}\label{app:qualitative}

\subsection{Example of Code Tasks and Watermarked Codes}\label{app:code-tasks}

\definecolor{codebg}{HTML}{FAFAFA}       %
\definecolor{codeframe}{HTML}{E0E0E0}    %
\definecolor{keyword}{HTML}{0000FF}      %
\definecolor{string}{HTML}{A31515}       %
\definecolor{comment}{HTML}{008000}      %
\definecolor{number}{HTML}{098658}       %
\definecolor{darklabel}{HTML}{404040}    %

\lstdefinestyle{prettypython}{
    language=Python,
    basicstyle=\ttfamily\scriptsize,
    commentstyle=\color{comment}\itshape,
    keywordstyle=\color{keyword}\bfseries,
    stringstyle=\color{string},
    numberstyle=\color{number},
    breaklines=true,
    breakatwhitespace=false,
    keepspaces=true,
    showspaces=false,
    showstringspaces=false,
    showtabs=false,
    tabsize=4,
    frame=none, %
    aboveskip=0pt,
    belowskip=0pt,
    columns=fullflexible, %
}
\lstset{style=prettypython}

\tcbset{
    enhanced,
    colback=codebg,
    colframe=codeframe,
    boxrule=1pt,
    arc=4pt,       %
    outer arc=4pt,
    boxsep=3pt,    %
    left=6pt, right=6pt, top=4pt, bottom=4pt, %
    drop shadow=black!20!white, %
    fonttitle=\bfseries\sffamily,
}

\begin{figure}[h!]
    \centering
    \begin{minipage}[t]{0.48\linewidth}
        {\sffamily\textbf{Input (Prompt + Canonical Solution):}} \vspace{0.5em}
        
        \begin{tcolorbox}[remember as=prompt]
\begin{lstlisting}
from typing import List

def has_close_elements(numbers: List[float], threshold: float) -> bool:
    """ Check if in given list of numbers, are any two numbers closer to each other than given threshold.
    >>> has_close_elements([1.0, 2.0, 3.0], 0.5)
    False
    >>> has_close_elements([1.0, 2.0, 5.0, 2.0], 0.3)
    True
    """
\end{lstlisting}
        \end{tcolorbox}
        
        \vspace{0.05em}
        
        \begin{tcolorbox}[remember as=solution]
\begin{lstlisting}
    for idx, elem in enumerate(numbers):
        for idx2, elem2 in enumerate(numbers):
            if idx != idx2:
                distance = abs(elem - elem2)
                if distance < threshold:
                    return True
    return False
\end{lstlisting}
        \end{tcolorbox}
    \end{minipage}
    \hfill
    \begin{minipage}[t]{0.48\linewidth}
        {\sffamily\textbf{Test:}} \vspace{0.5em}
        
        \begin{tcolorbox}[remember as=test]
\begin{lstlisting}
METADATA = { 'author': 'jt', 'dataset': 'test' }

def check(candidate):
    assert candidate([1.0, 2.0, 3.9, 4.0, 5.0, 2], 0.3) == True
    assert candidate([1.0, 2, 3, 4.0, 5.0, 2.2], 0.05) == False
    assert candidate([1.0, 2.0, 5.9, 4.0, 5.0, 0.95]) == True
    assert candidate([1.0, 2.0, 5.9, 4.0, 5.0, 0.8]) == False
    assert candidate([1.0, 2.0, 3.0, 4.0, 5.0, 2], 0.1) == True

check(has_close_elements)
\end{lstlisting}
        \end{tcolorbox}
    \end{minipage}
    
    \vspace{1em} %
    \caption{Example of a HumanEval task. 
    The left block represents the text subjected to post-hoc watermarking (paraphrasing). 
    The right block is the test harness used to verify functional correctness (Pass@1).}
    \label{fig:humaneval_example}
\end{figure}

\newcolumntype{Y}{>{\raggedright\arraybackslash}X}

\begin{xltabular}{\linewidth}{@{}YY@{}}
    \caption{Post-hoc code watermarking examples on a sample from MBPP (using Llama-3.1-8B at temperatures 0.8, 1.2 and 1.4 with Gumbel-max watermarking, top-$p=0.95$).} 
    \label{tab:more_code_example} \\
    
    \toprule
    \textbf{Original Code} & \textbf{Watermarked Code} \\
    \midrule
    \endfirsthead
    
    \caption[]{Post-hoc code watermarking at different temperatures (continued)} \\
    \toprule
    \textbf{Original Code} & \textbf{Watermarked Code} \\
    \midrule
    \endhead
    
    \endfoot
    
    \bottomrule
    \endlastfoot

    \ttfamily\scriptsize
    \# A function to find the minimum cost path to reach (m, n) from (0, 0) ... \newline
    R = 3\ \newline
    C = 3\ \newline
    def min\_cost(cost, m, n): \ \newline
    \hspace*{1em}tc = [[0 for x in range(C)] for x in range(R)] \ \newline
    \hspace*{1em}tc[0][0] = cost[0][0] \ \newline
    \hspace*{1em}for i in range(1, m+1): \ \newline
    \hspace*{1em}\hspace*{1em}tc[i][0] = tc[i-1][0] + cost[i][0] \ \newline
    \hspace*{1em}for j in range(1, n+1): \ \newline
    \hspace*{1em}\hspace*{1em}tc[0][j] = tc[0][j-1] + cost[0][j] \ \newline
    \hspace*{1em}for i in range(1, m+1): \ \newline
    \hspace*{1em}\hspace*{1em}for j in range(1, n+1): \ \newline
    \hspace*{1em}\hspace*{1em}\hspace*{1em}tc[i][j] = min(tc[i-1][j-1], tc[i-1][j], tc[i][j-1]) + cost[i][j] \ \newline
    \hspace*{1em}return tc[m][n] \newline
    
    & 
    \ttfamily\scriptsize
    \# A function to calculate the minimum cost path from the origin (0, 0) ... \newline
    R = 3 \newline
    C = 3 \newline
     \newline
    def min\_cost\_path(cost, m, n): \newline
    \hspace*{1em}\# Create a temporary cost matrix tc of the same dimensions... \newline
    \hspace*{1em}tc = [[0 for x in range(C)] for x in range(R)] \newline
    \hspace*{1em}\# The cost to reach the first cell (0, 0) is the cost... \newline
    \hspace*{1em}tc[0][0] = cost[0][0] \newline
    \hspace*{1em}\# Calculate the cost to reach the first row. \newline
    \hspace*{1em}for i in range(1, m+1): \newline
    \hspace*{2em}\# The cost to reach the first cell in the current row... \newline
    \hspace*{2em}tc[i][0] = tc[i-1][0] + cost[i][0] \newline
    \hspace*{1em}\# Calculate the cost to reach the first column. \newline
    \hspace*{1em}for j in range(1, n+1): \newline
    \hspace*{2em}\# The cost to reach the first cell in the current column... \newline
    \hspace*{2em}tc[0][j] = tc[0][j-1] + cost[0][j] \newline
    \hspace*{1em}\# Calculate the cost to reach the rest of the cells. \newline
    \hspace*{1em}for i in range(1, m+1): \newline
    \hspace*{2em}for j in range(1, n+1): \newline
    \hspace*{3em}\# The cost to reach the current cell is the minimum... \newline
    \hspace*{3em}tc[i][j] = min(tc[i-1][j-1], tc[i-1][j], tc[i][j-1]) + cost[i][j] \newline
    \hspace*{1em}\# Return the cost to reach the cell at position (m, n). \newline
    \hspace*{1em}return tc[m][n] \\
    \midrule
    \multicolumn{2}{c}{\footnotesize \textbf{Stats:} $T$: $0.8$ \statsep $p_{\text{value}}$: $2.11e-01$ \statsep Tokens: $414$ \statsep Code correctness: True} \\
    \midrule
    
    \ttfamily\scriptsize
    \# A function to find the minimum cost path to reach (m, n) from (0, 0) ... \newline
    R = 3\ \newline
    C = 3\ \newline
    def min\_cost(cost, m, n): \ \newline
    \hspace*{1em}tc = [[0 for x in range(C)] for x in range(R)] \ \newline
    \hspace*{1em}tc[0][0] = cost[0][0] \ \newline
    \hspace*{1em}for i in range(1, m+1): \ \newline
    \hspace*{1em}\hspace*{1em}tc[i][0] = tc[i-1][0] + cost[i][0] \ \newline
    \hspace*{1em}for j in range(1, n+1): \ \newline
    \hspace*{1em}\hspace*{1em}tc[0][j] = tc[0][j-1] + cost[0][j] \ \newline
    \hspace*{1em}for i in range(1, m+1): \ \newline
    \hspace*{1em}\hspace*{1em}for j in range(1, n+1): \ \newline
    \hspace*{1em}\hspace*{1em}\hspace*{1em}tc[i][j] = min(tc[i-1][j-1], tc[i-1][j], tc[i][j-1]) + cost[i][j] \ \newline
    \hspace*{1em}return tc[m][n] \newline
    
    & 
    \ttfamily\scriptsize
    \# Function to calculate the minimum cost path from the origin (0, 0) ... \newline
    R, C = 3, 3 \newline
     \newline
    def min\_cost\_path(cost, m, n): \newline
    \hspace*{1em}dp = [[0 for \_ in range(C)] for \_ in range(R)] \newline
    \hspace*{1em}dp[0][0] = cost[0][0] \newline
     \newline
    \hspace*{1em}\# Calculate the minimum cost to reach the first row. \newline
    \hspace*{1em}for i in range(1, m+1): \newline
    \hspace*{2em}dp[i][0] = dp[i-1][0] + cost[i][0] \newline
     \newline
    \hspace*{1em}\# Calculate the minimum cost to reach the first column. \newline
    \hspace*{1em}for j in range(1, n+1): \newline
    \hspace*{2em}dp[0][j] = dp[0][j-1] + cost[0][j] \newline
     \newline
    \hspace*{1em}\# Calculate the minimum cost to reach the rest of the positions. \newline
    \hspace*{1em}for i in range(1, m+1): \newline
    \hspace*{2em}for j in range(1, n+1): \newline
    \hspace*{3em}dp[i][j] = min(dp[i-1][j-1], dp[i-1][j], dp[i][j-1]) + cost[i][j] \newline
     \newline
    \hspace*{1em}return dp[m][n] \\
    \midrule
    \multicolumn{2}{c}{\footnotesize \textbf{Stats:} $T$: $1.2$ \statsep $p_{\text{value}}$: $2.63e-02$ \statsep Tokens: $256$ \statsep Code correctness: True} \\
    \midrule

    \ttfamily\scriptsize
    \# A function to find the minimum cost path to reach (m, n) from (0, 0) ... \newline
    R = 3\ \newline
    C = 3\ \newline
    def min\_cost(cost, m, n): \ \newline
    \hspace*{1em}tc = [[0 for x in range(C)] for x in range(R)] \ \newline
    \hspace*{1em}tc[0][0] = cost[0][0] \ \newline
    \hspace*{1em}for i in range(1, m+1): \ \newline
    \hspace*{1em}\hspace*{1em}tc[i][0] = tc[i-1][0] + cost[i][0] \ \newline
    \hspace*{1em}for j in range(1, n+1): \ \newline
    \hspace*{1em}\hspace*{1em}tc[0][j] = tc[0][j-1] + cost[0][j] \ \newline
    \hspace*{1em}for i in range(1, m+1): \ \newline
    \hspace*{1em}\hspace*{1em}for j in range(1, n+1): \ \newline
    \hspace*{3em}tc[i][j] = min(tc[i-1][j-1], tc[i-1][j], tc[i][j-1]) + cost[i][j] \ \newline
    \hspace*{1em}return tc[m][n] \newline
    
    & 
    \ttfamily\scriptsize
    \# Function to calculate the minimum cost path from the origin (0, 0) ... \newline
    RatingsColumns = 3 \newline
    Rows = 3 \newline
     \newline
    def min\_expense(cost, m, n): \newline
    \hspace*{1em}databaseOfDiscountedValues = [[0 for xPos in range(RatingsColumns)] for yPos in range(Rows)]  \newline
    \hspace*{1em}initialPositionValuation = cost[0][0]  \newline
    \hspace*{1em}for tabIndex in range(1, Rows+1):  \newline
    \hspace*{1em}\hspace*{1em}databaseOfDiscountedValues[tabIndex][0] = databaseOfDiscountedValues[tabIndex-1][0] + cost[tabIndex][0]  \newline
    \hspace*{1em}for tabIndex in range(1, RatingsColumns+1):  \newline
    \hspace*{1em}\hspace*{1em}databaseOfDiscountedValues[0][tabIndex] = databaseOfDiscountedValues[0][tabIndex-1] + cost[0][tabIndex]  \newline
    \hspace*{1em}for tabIndex in range(1, Rows+1):  \newline
    \hspace*{1em}\hspace*{1em}for colIndex in range(1, RatingsColumns+1):  \newline
    \hspace*{3em}dbDV = min(databaseOfDiscountedValues[tabIndex-1][colIndex-1], databaseOfDiscountedValues[tabIndex-1][colIndex], databaseOfDiscountedValues[tabIndex][colIndex-1]) + cost[tabIndex][colIndex]  \newline
    \hspace*{3em}databaseOfDiscountedValues[tabIndex][colIndex] = dbDV  \newline
    \hspace*{1em}return databaseOfDiscountedValues[m][n] \\
    \midrule
    \multicolumn{2}{c}{\footnotesize \textbf{Stats:} $T$: $1.4$ \statsep $p_{\text{value}}$: $8.14e-11$ \statsep Tokens: $302$ \statsep Code correctness: False} \\
    
\end{xltabular}

\newpage
\subsection{Descriptive Text and Dialogue Preservation}

\begin{table}[h!]
    \centering
    \caption{\textbf{Additional qualitative comparisons on literary text.} 
    Additional samples from \textit{A Christmas Carol} by Charles Dickens, rephrased using Llama-3.2-3B-Instruct with the Gumbel-max scheme, $T=0.7$ and top-p = 0.95. 
    All outputs maintain strong semantic similarity and watermark detectability.}
    \small
    \begin{tabularx}{\linewidth}{@{}YY@{}}
        \toprule
        \textbf{Original Text} & \textbf{Watermarked Rephrasing} \\
        \midrule
        \multicolumn{2}{c}{\textit{Example 1}} \\
        \midrule
Lumber-room as usual. Old fire-guard, old shoes, two fish-baskets, washing-stand on three legs, and a poker.\newline
Quite satisfied, he closed his door, and locked himself in; double-locked himself in, which was not his custom. Thus secured against surprise, he took off his cravat; put on his dressing-gown and slippers, and his nightcap; and sat down before the fire to take his gruel.\newline
It was a very low fire indeed; nothing on such a bitter night. He was obliged to sit close to it, and brood over it, before he could extract the least sensation of warmth from such a handful of fuel. The fireplace was an old one, built by some Dutch merchant long ago, and paved all round with quaint Dutch tiles, designed to illustrate the Scriptures. There were Cains and Abels, Pharaoh's daughters; Queens of Sheba, Angelic messengers descending through the air on clouds like feather-beds, Abrahams, Belshazzars, Apostles putting off to sea in butter-boats, hundreds of figures to attract his thoughts; and yet that face of Marley, seven years dead, came like the ancient Prophet's rod, and swallowed up the whole.
&
Lumber-room as usual. Old fire-guard, worn shoes, two fish-baskets, a washing-stand on three legs, and a poker. He was quite content, so he locked himself in, double-locked himself in, which wasn't his usual practice. Thus secure against surprise, he shed his cravat, put on his dressing-gown, slippers, and his nightcap; and sat down before the fire to enjoy his gruel.\newline
It was a very low fire indeed; the flames seemed to be burning with a fierce intensity. He was forced to sit close to it, and reflect on it, before he could feel any warmth emanating from such a small amount of fuel. The fireplace was an old one, built by a Dutch merchant many years ago, and adorned with intricate tiles depicting biblical scenes. There were numerous figures from the Bible, including Cain and Abel, Pharaoh's daughters, the Queen of Sheba, and the Angelic messengers descending from the heavens like angels on cloud-like bodies; yet, his gaze fell upon the face of Marley, seven years deceased, and it seemed to hold a profound power, as if the very spirit of the departed was still present.
\\
        \midrule
        \multicolumn{2}{c}{\footnotesize \textbf{Stats:} SBERT: 0.925 \statsep Ratio: 1.03  \statsep Perplexity: 3.10 \statsep $p$-val: $6.7 \times 10^{-4}$} \\
        \midrule
        \multicolumn{2}{c}{\textit{Example 2}} \\
        \midrule
`` again; and followed it up with ``Humbug.''\newline
``Don't be cross, uncle!'' said the nephew.\newline
``What else can I be,'' returned the uncle, ``when I live in such a world of fools as this? Merry Christmas! Out upon merry Christmas! What's Christmas time to you but a time for paying bills without money; a time for finding yourself a year older, but not an hour richer; a time for balancing your books and having every item in 'em through a round dozen of months presented dead against you? If I could work my will,'' said Scrooge indignantly, ``every idiot who goes about with 'Merry Christmas' on his lips, should be boiled with his own pudding, and buried with a stake of holly through his heart. He should!''\newline
``Uncle!'' pleaded the nephew.\newline
``Nephew!'' returned the uncle sternly, ``keep Christmas in your own way, and let me keep it in mine.
&
``Again; and it followed with an exclamation, ``Humbug!'' \newline
``Don't be cross, uncle!'' said the nephew.\newline
``What else can I be,'' returned the uncle, ``when I live in such a world of fools as this? Christmas is a time when everyone is trying to deceive themselves, pretending to be merry and happy, but in reality, it's just a chore to pay bills and get older without earning more. It's a time when people make lists and check them twice, but never actually achieve anything. If I could change my ways,'' said Scrooge indignantly, ``every fool who goes about spreading 'Merry Christmas' on their lips should be shunned and left to rot. They should be boiled in their own misery, and buried with a curse of isolation through the desolate winter months.''\newline
``Uncle!'' pleaded the nephew.\newline
``Nephew!'' returned the uncle sternly, ``keep Christmas in your own way, and let me keep it in mine.''
\\
        \midrule
        \multicolumn{2}{c}{\footnotesize \textbf{Stats:} SBERT: 0.935 \statsep Ratio: 1.04  \statsep Perplexity: 3.28 \statsep $p$-val: $2.7 \times 10^{-5}$} \\
        \bottomrule
    \end{tabularx}
    \label{tab:appendix_examples}
\end{table}

\newpage

\section{Additional Results}\label{app:additional_results}

\subsection{Analysis of Output Length Distributions}
\label{app:length_distrib}

In the main text, we apply a filtering criterion to retain only rephrasings where the output length remains within a factor of $0.75$ to $1.25$ of the original input length. 
In this appendix, we analyze the distribution of these length ratios to validate that this filtering does not introduce bias against specific watermarking schemes.

\begin{figure*}[b!]
    \centering
    \includegraphics[width=\textwidth]{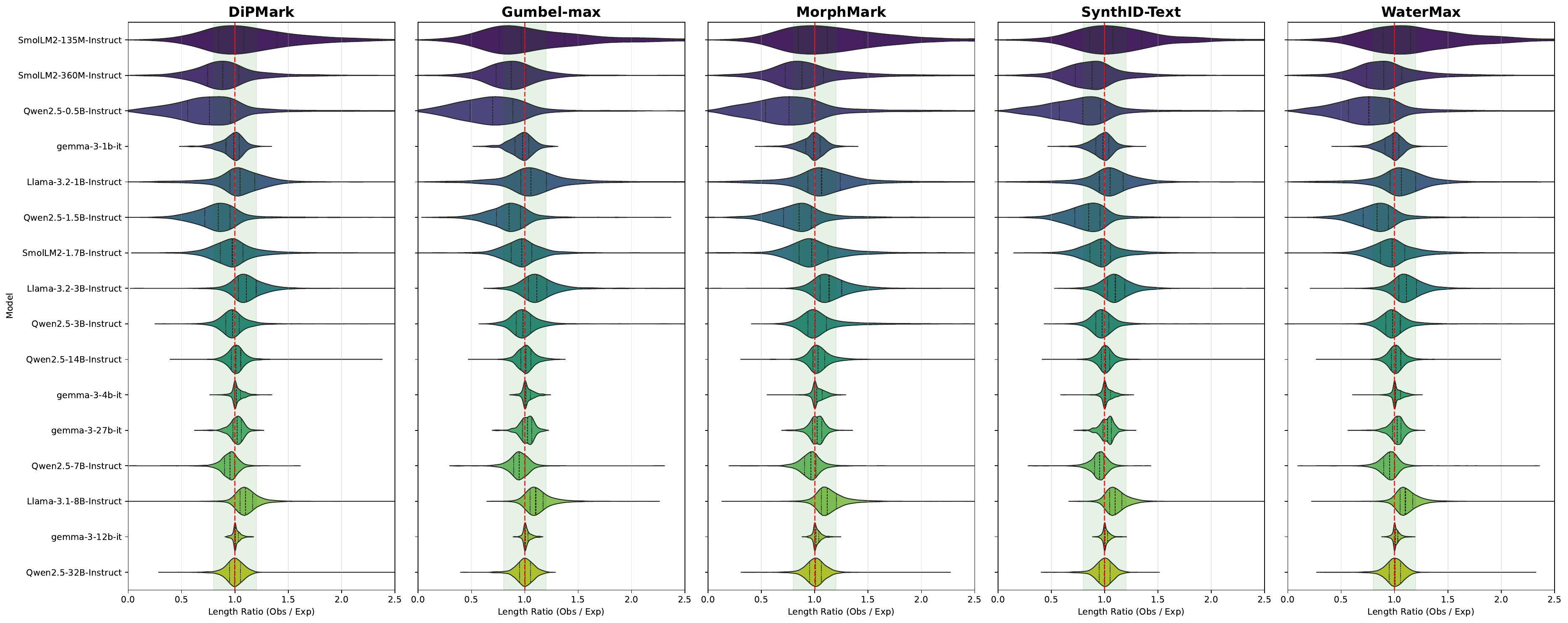} %
    \caption{\textbf{Impact of Model Choice on Output Length Consistency.} 
    Violin plots showing the distribution of length ratios ($\text{Length}_{\text{wm}} / \text{Length}_{\text{orig}}$) for varying models and watermarking schemes. 
    The green shaded region represents the inclusion criteria ($[0.75, 1.25]$) used in the main experiments. The red dashed line indicates the ideal ratio of $1.0$. Note that length variance is driven primarily by the base model choice (rows) rather than the watermarking scheme (columns), with larger models consistently adhering closer to the target length.}
    \label{fig:length_distribs}
\end{figure*}

\autoref{fig:length_distribs} presents the distribution of length ratios (defined as $L_{\text{output}} / L_{\text{input}}$) across all evaluated models and five distinct watermarking methods. The green shaded region indicates the acceptance window used in our main experiments.

We observe two key trends:
\begin{enumerate}
    \item \textbf{Consistency across schemes:} For any given base model (rows), the distribution of output lengths is remarkably consistent across all watermarking methods (columns). Whether using simple rejection sampling (WaterMax) or distribution perturbation (DiPMark, Gumbel-max), the variance in output length remains stable.
    \item \textbf{Dependence on model capability:} The ability to respect the length constraint is primarily a function of the model size and instruction-following capability. Smaller models (e.g., SmolLM2-135M) exhibit high variance and frequently generate outputs that are too short or too long, whereas larger, more capable models (e.g., Llama-3.1-8B, Gemma-2-27B) produce tight distributions centered near the ideal ratio of $1.0$.
\end{enumerate}

These findings confirm that outliers in length are attributable to the underlying model's generation stability rather than the watermarking process itself.
Consequently, filtering these outliers allows for a fairer assessment of semantic preservation and detection power on valid generations, rather than penalizing watermarking schemes for the base model's verbosity or brevity failures.

\end{document}